\documentclass[fleqn,twoside]{article}

\usepackage[headings]{espcrc2}

\usepackage{graphicx,epsfig, amsmath}

\newcommand{\tr} {\rm tr}
\newcommand {\CalA} {\mathcal A}
\newcommand {\CalB} {\mathcal B}
\newcommand {\CalC} {\mathcal C}
\newcommand {\CalD} {\mathcal D}

\newcommand {\CalF} {\mathcal F}

\newcommand {\CalH} {\mathcal H}
\newcommand {\CalI} {\mathcal I}

\newcommand {\CalO} {\mathcal O}
\newcommand {\CalN} {\mathcal N}
\newcommand {\CalL} {\mathcal L}
\newcommand {\CalS} {\mathcal S}

\newcommand {\CalM} {\mathcal M}

\newcommand {\CalU} {\mathcal U}

\newcommand {\CalV} {\mathcal V}
\newcommand {\CalW} {\mathcal W}
\newcommand {\CalY} {\mathcal Y}
\newcommand {\CalZ} {\mathcal Z}
\newcommand {\ba} {\mathbf a}
\newcommand {\bb} {\mathbf b}
\newcommand {\bH}   {\mathbf H}
\newcommand {\bR}   {\mathbf R}
\newcommand {\bZ}   {\mathbf Z}
\newcommand {\bC}   {\mathbf C}

\newcommand {\CP}   {\mathbf C \mathbf P}

\newcommand{\gm} {\rm g}
\newcommand{\hm} {\rm h}
\newcommand {\al} {\alpha}

\newcommand {\be} {\beta}

\newcommand {\de} {\delta}
\newcommand {\ve}  {\varepsilon}
\newcommand {\ep}  {\epsilon}
\newcommand {\ga} {\gamma}

\newcommand {\lam}  {\lambda}
\newcommand {\n} {\nu}
\newcommand {\m} {\mu}
\newcommand {\si}   {\sigma}

\newcommand {\om} {\omega}

\newcommand {\p} {\partial}

\title{Darboux coordinates, Yang-Yang functional, and gauge theory}

\author{N. Nekrasov\address[IHES]{IHES, 35 route de Chartres\\ 
        91440 Bures-sur-Yvette, FRANCE}$^{,}$\address[ITEP]{ITEP, Bol. Cheremushkinskaya 25\\
        117259 Moscow, RUSSIA}\thanks{On leave of absence}$^{,}$\address[SCGP]{Simons Center for Geometry and Physics, Stony Brook University, Stony Brook NY 11794 USA}\thanks{On leave of absence},
        A. Rosly\addressmark[ITEP]$^{,}$\address[LAG]{Laboratory of Algebraic Geometry, SU-HSE, 7 Vavilova Str., \\
117312 Moscow, RUSSIA} and
        S. Shatashvili\addressmark[IHES]\thanks{IHES Louis Michel Chair}\address{Hamilton Mathematics Institute, Trinity College Dublin \\
        Dublin 2, IRELAND}$^{,}$\address{School of Mathematics, Trinity College Dublin \\
        Dublin 2, IRELAND}
        }

\runtitle{Darboux, Yang-Yang, and Yang-Mills}
\runauthor{Nekrasov, Rosly, Shatashvili}

\begin{document}

\begin{abstract}
The moduli space of $SL_2$ flat connections on a punctured Riemann surface $\Sigma$ with the fixed conjugacy classes of the monodromies around the punctures is endowed with a system of holomorphic Darboux coordinates, in which the generating function of the variety of $SL_{2}$-opers
is identified with the universal part of the effective twisted superpotential of the corresponding four dimensional ${\CalN} = 2$ supersymmetric theory 
subject to the two-dimensional $\Omega$-deformation. This allows to give a definition of the Yang-Yang functionals for the quantum Hitchin system in terms of the classical geometry of the moduli space of local systems for the dual gauge group, and connect it to the instanton counting of the four dimensional gauge theories, in the rank one case. 

\bigskip
\bigskip
\noindent
TCD MATH 11-05\\
\noindent
HMI-11-04\\
\noindent
IHES/P/11/16\\
\noindent
ITEP-TH-16/11\\

\vspace{1pc}
\end{abstract}

\maketitle

\section{Introduction}

The exact relation between the microscopic definition of a quantum field theory and its low energy behavior is the major research subject.
In the context of the supersymmetric gauge theories in four and two dimensions this relation touches upon some unexpected domains of the mathematical physics, such as the theory of classical and quantum integrable systems, and, in particular, the celebrated Bethe ansatz.  

\subsection{Bethe ansatz}

Bethe ansatz is a useful technique for finding the spectra of the quantum integrable systems, such as the spin chains or the many-body systems, or even the 1+1 dimensional quantum field theories, see \cite{Faddeev:1993hd}. 

Generally speaking, the ansatz consists in finding a set of states ${\bf\Psi}({\lam}_{1}, \ldots , {\lam}_{N})$ which are characterized using some algebraic structure, or in terms of some functional equations, or otherwise. The physical meaning of the parameters ${\lam}_{1}, \ldots, {\lam}_{N}$ may differ from context to context, yet often they are the quasi-momenta of the quasi-free constituents. 

 The condition that ${\bf\Psi}({\lam}_{1}, \ldots , {\lam}_{N})$ actually belongs to the (Hilbert) space of states of the model, and it is the joint eigenstate of all the commuting Hamiltonians translates to the set of $N$ equations on the quasimomenta ${\lam}_{1}, \ldots , {\lam}_{N}$ which, remarkably, have the potential:
\begin{equation}
\frac{{\p}Y({\lam})}{{\p}{\lam}_{k}} = 2\pi i n_{k}, \qquad  k = 1, \ldots N
\label{eq:yyfun}
\end{equation}
The function $Y({\lam})$ is often analytic multi-valued. We call it the Yang-Yang function, since C.N~Yang and C.P.~Yang used it for the analysis of the non-linear Schr\"odinger model \cite{Yang:1968rm} (see also 
\cite{Yang:1966ty}). 

As an illustration, consider the simplest spin chain, the $SU(2)$
Heisenberg magnet, with $L$ spin sites. It can be solved with the help
of the algebraic Bethe ansatz, where the eigenstate
of all the commuting Hamiltonians is found in the
form of the $N$ "quasiparticles"
\begin{equation}
\label{eq:bethst}
B({\lam}_{1})B({\lam}_{2})\ldots B({\lam}_{N})
\vert {\rm vac} \rangle
\end{equation}
where the quasimomenta ${\lam}_{k}$ solve the equation
(\ref{eq:yyfun}) 
with the function
\begin{equation}
\label{eq:yhs}
Y({\lam}) = \sum_{i=1}^{N} L {\varpi}_{1\over 2}
({\lam}_{i}) 
 + \sum_{i<j}^{N} {\varpi}_{1} ({\lam}_{i}- {\lam}_{j})
 \end{equation}
 where ${\varpi}_{s}({\lam})$ is a certain special function whose explicit form is known:
$$
{\varpi}_{s}({\lam}) = \qquad\qquad\qquad\qquad\qquad\qquad\qquad\qquad
$$
\begin{equation}
\label{eq:vpi}
\ \left( {\lam} + is \right) {\log} \left( {\lam} + is \right)
- \left( {\lam} - is \right) {\log} \left( {\lam} - is \right) 
\end{equation}
and $B(\lam)$ is the  creation operator (see \cite{Faddeev:1979gh} for details) of the quasiparticle of quasimomentum ${\lam}$. 
The eigenvalues of all the commuting Hamiltonians are written in terms of the solutions to (\ref{eq:yyfun}). 
In addition, the function $Y({\lam})$, its derivatives with respect to the parameters, or its Hessian enter in the explicit expressions for the norms of the Bethe vectors (\ref{eq:bethst}), the correlation functions of local operators (see e. g.  \cite{Fedya} and the references there). 
Thus the function $Y(\lam)$ plays a central role in the concise formulation of the solution of the quantum system. 

{}The universality of (\ref{eq:yyfun}) is so far an experimental fact about the world of the quantum integrability.
It has been a somewhat puzzling (and therefore for a long time abandonded) question as to what is the meaning of the function $Y({\lam})$, why is the spectrum of the quantum problem described by what looks like a classical equation (\ref{eq:yyfun})? why does the function $Y({\lam})$ look like a classical action of some classical mechanical system? 

The goal of this paper is to elucidate this point. We shall show (in the restricted class of integrable systems) that
one can indeed associate to the quantum integrable system a classical mechanical model, i.e. a symplectic manifold and a Lagrangian submanifold, and that the function $Y({\lam})$ is identified with the generating function of this submanifold, in the appropriate Darboux coordinates. Thus we give a geometric definition of the Yang-Yang function
for a large class of quantum integrable systems.

\subsection{Gauge theories and quantum integrability}

It turns out that quantum gauge theories are the way to understand the correspondence between the quantum integrability and classical symplectic geometry.  

 In \cite{Moore:1997dj} a connection between a quantum integrable system, the $N$-particle sector of the non-linear Schr\"odinger theory, and a topologically twisted two dimensional gauge theory, was observed. The coupling constant of the quantum integrable system maps to the equivariant parameter, the twisted mass of the gauge theory, i.e. to the bulk coupling.  
In \cite{Gerasimov:2006zt}, \cite{Gerasimov:2007ap} this subject was revived, by showing that the observation
of  \cite{Moore:1997dj} can be interpreted as the statement that the $2$-observable \cite{Witten:1992} of the topological gauge theory of \cite{Moore:1997dj} descends from the Yang-Yang function \cite{Yang:1968rm} of the system of 
$N$ non-relativistic particles on a circle with the ${\delta}$-function pairwise potential  (equivalent to the $N$-particle sector of the non-linear Schr\"odinger system). Together with the earlier work on the relation between the quantum integrable many-body and spin systems of the Calogero-Moser-Sutherland type, harmonic analysis, and the topological gauge theories in various spacetime dimensions \cite{Gorsky:1994ec},  \cite{Gorsky:1993dq}, \cite{Gorsky:1994dj}, \cite{Gorsky:1993pe} (where the strength of the interaction on the  quantum system side corresponds to the parameters of the line defects on the gauge theory side) this strongly suggested that the connection 
between the gauge theory, the representation theory, and the quantum integrable systems is  universal. 

In all these cases the spectrum of the observables in gauge theory maps to the spectrum of quantum Hamiltonians on the integrable theory side.

Eventually in \cite{Nekrasov:2009uh}, \cite{Nekrasov:2009zz}, \cite{Nekrasov:2009ui} the precise form of the Bethe/Gauge correspondence was formulated:

{\em The supersymmetric vacua of the gauge theories with the two dimensional ${\CalN}=2$ super-Poincare invariance, with the generic twisted masses and the superpotential, are the stationary states (the common eigenstates of the commuting Hamiltonians) of some quantum integrable system. The commuting Hamiltonians are the generators of the twisted chiral ring of the gauge theory.
The quasimomenta of Bethe particles are the special coordinates on quasiclassical moduli space of vacua (the genericity assumption on the masses and superpotential implies this is a Coulomb branch). The Yang-Yang functional, generating the Bethe equations of the quantum system, is the effective twisted superpotential of the gauge theory. Thus, Bethe equations single out the supersymmetric vacua of the gauge theory.}

This correspondence was checked for
a large class of spin systems including the {\sl XXX, XXZ, XYZ} spin chains for all spin groups and impurities, and for some quantum algebraic integrable
systems as the elliptic Calogero-Moser, its relativistic version and limits such as the periodic Toda chain.

The gauge theories with the two dimensional super-Poincare invariance need not be two dimensional. In fact, in \cite{Nekrasov:2009rc} the four dimensional theories with four dimensional super-Poincare invariance subject to the $\Omega$-deformation in two dimensions were studied, leading to the quantum integrable systems whose classical limits are the Seiberg-Witten integrable systems \cite{Gorsky:1995zq}, \cite{Martinec:1995}, \cite{Donagi:1995cf}, \cite{Donagi:1997sr}  describing the moduli space of vacua of the original four dimensional ${\CalN}=2$ supersymmetric theory. The $\Omega$-background in quantum field theory was introduced in \cite{Nekrasov:2002qd}, the idea is based on the earlier work \cite{Lossev:1997bz}, \cite{Losev:1997tp}. Its partition function, ${\CalZ}$-function, plays an important role in what follows. 
 
\subsection{Gauge theories from six dimensions}

This paper will study a limited set of four dimensional 
${\CalN}=2$ gauge theories, namely those which can be 
 engineered by taking the six dimensional (0,2) superconformal theory
(discovered in \cite{Witten:1995zh}, \cite{Strominger:1995ac})  and compactifying it with a partial twist on a Riemann surface \cite{Witten:1997sc}, \cite{Gaiotto:2009we}, \cite{Gaiotto:2009hg}. These theories have the ADE classification, and in what follows we shall denote them by their $ADE$ type.

The resulting theory can be analyzed in several ways. 
One the one hand, by assuming the size of $\Sigma$ negligible one deals with the four dimensional superconformal theory with ${\CalN}=2$ supersymmetry. The enhancement of the ${\CalN}=2$ supersymmetry to the superconformal symmetry is clear from the superconformal symmetry of the six dimensional $(0,2)$ theory. 

Indeed, consider first the compactification of the six dimensional theory on a finite size Riemann surface $\Sigma$, with the metric $g_{2}$. Let the metric on the four dimensional spacetime $X^4$ be $g_{4}$. The Lagrangian of the theory on $X^4$ depends on $g_2$. 

If the six dimensional theory were a gauge theory to begin with, then the resulting four dimensional theory would have had the gauge coupling ${\bf g}_{6}$ depending on the symplectic, i.e. K\"ahler, moduli of $\Sigma$, e.g.
${\bf g}_{4}^{-2} = {\bf g}_{6}^{-2} {\rm Area}_{\Sigma}$, where 
\begin{equation}
\label{eq:ar}
{\rm Area}_{\Sigma} = \int_{\Sigma} \sqrt{{\rm det}(g_{2})}
\end{equation}
However the $(0,2)$ theory in six dimensions is not a gauge theory, and the relation between the couplings in four dimensions and the geometry of $\Sigma$ is subtle. 
To begin with, one gets not one, but several gauge group factors, depending on the topology of $\Sigma$, and their couplings remain finite even when ${\rm Area}_{\Sigma} \to 0$. The couplings are determined by the complex structure of $\Sigma$, determined by the conformal class $[g_{2}]$ of $g_{2}$. 

The conformal transformation of the four dimensional metric on $X^4$
\begin{equation}
\label{ref:dsf}
g_{4} \mapsto {\l}^{2} g_{4}
\end{equation}
lifts to the conformal transformation of the six dimensional metric
\begin{equation}
\label{eq:dss}
g_{6} = g_{2} \oplus g_{4}  \mapsto {\l}^{2} g_{6} =
{\l}^{2} g_{2} \oplus {\l}^2 g_{4} 
\end{equation}
on
${\Sigma} \times X^4$. If $\Sigma$ has a finite size metric, then the resulting six dimensional metric in the right hand side of (\ref{eq:dss}) gives rise to a different metric on $\Sigma$. However, in the limit of vanishing area of $\Sigma$ the difference is negligible, hence the theory, in this limit, becomes conformal in the four dimensional sense.  

It is not trivial to identify what this theory is, in four dimensional terms, e.g. describe the matter content and the Lagrangian. 

When $\Sigma$ is a two-torus, one can use the string duality 
dictionary to conclude that this theory is the ${\CalN}=4$ super-Yang-Mills theory, whose gauge group is determined by the type of the six dimensional superconformal theory we started with. In particular, for the $A_{1}$ theory in six dimensions we get the $SU(2)$ theory in four dimensions. 
When $\Sigma$ is a genus $g$ Riemann surface, Gaiotto argues \cite{Gaiotto:2009we} one gets the ${\CalN}=2$ super-Yang-Mills with the gauge group $SU(2)^{3g-3}$ with $2g-2$ hypermultiplets transforming in the 
tri-fundamental representations of some triples of $SU(2)$'s out of the total $3g-3$ factors. 
The precise assignment of these triplets is not unique, it is encoded in the trivalent graph describing the maximal degeneration of the complex structure of 
$\Sigma$. In particular, the complexified gauge couplings of the $SU(2)$ groups are identified with the usual asymptotic complex structure moduli corresponding to the pinched handles on the degenerate Riemann surface. 
In \cite{Gaiotto:2009we} more general theories, corresponding to the genus $g$ complex curves with $n$ punctures and some local data, assigning a complex number ${\nu}_{k}$ to the puncture $z_{k}$, were proposed. In particular,
the celebrated ${\CalN}=2$ $SU(2)$ theory with $N_{f}=4$ fundamental hypermultiplets corresponds to the genus $0$ curve, a sphere, with $4$ punctures. The one-dimensional moduli space ${\overline{\CalM}}_{0,4} \approx {\CP}^{1}$ of complex structures
of the $4$-punctured sphere parametrizes the gauge coupling of the $SU(2)$ gauge group. Remarkably, this theory already exhibits the non-trivial S-duality of the ${\CalN}=2$ theory. There are three points of the maximal degeneration in 
${\overline{\CalM}}_{0,4}$. In the neighborhood of each point one identifies the gauge theory with the $SU(2)$ theory with 
four fundamental hypers, however, the relation between the gauge couplings and the matter multiplet masses in the three respective weak coupling regions is nontrivial -- these are not the same gauge groups and not the same matter multiplets, as has been already observed in \cite{Seiberg:1994aj} at the level of the Seiberg-Witten curves. In particular, the triality exchanging the representations of the global $SO(8)$ symmetry group is identified in \cite{Gaiotto:2009we} with the modular group of ${\CalM}_{0,4}$.

\subsection{The $\Omega$-background}
Any ${\CalN}=2$ supersymmetric theory in four dimensions can be subject to the $\Omega$-deformation. This is achieved in three steps: i) given the four dimensional theory $T_{4}$ find a six dimensional ${\CalN}=1$ supersymmetric gauge theory $T_{6}$, whose dimensional reduction yields $T_{4}$; ii) compactify $T_{6}$ on a manifold $X^{6}$ which is an ${\bR}^{4}$ vector bundle over the two-torus ${\bf T}^{2}$, of area $r^{2}$ with a flat $Spin(4) = SU(2)_{+} \times SU(2)_{-}$ connection, whose holonomies around the two non-contractible cycles are $( e^{\frac{ir}{2}  {\rm Re}({\ve}_{1} + {\ve}_{2}) {\si}_{3}}, e^{\frac{ir}{2}   {\rm Re}({\ve}_{1} - {\ve}_{2})  {\si}_{3} })$ and $( e^{\frac{ir}{2}  {\rm Im}({\ve}_{1} + {\ve}_{2})  {\si}_{3}}, e^{\frac{ir}{2}   {\rm Im}({\ve}_{1} - {\ve}_{2})  {\si}_{3}} )$, respectively. In addition, embed the $SU(2)_{+}$ part of the flat connection into the $R$-symmetry $SU(2)$ of the six dimensional theory; iii) take the limit $r \to 0$ while keeping the complex numbers ${\ve}_{1}, {\ve}_{2}$ finite. In this way we get the $\Omega$-deformed theory on ${\bR}^{4}$. The parameters ${\ve}_{1}, {\ve}_{2}$ first appeared as the equivariant parameters in the integrals over the instanton and D-brane moduli spaces in \cite{Moore:1997dj}, \cite{Moore:1998et}. 

The embedding of $SU(2)_{+}$ into the $R$-symmetry group is not unique when there are matter multiplets, as one can also embed $SU(2)_{+}$ in the global symmetry group. In other words, the masses of the four dimensional matter multiplets can be shifted by the multiples of ${\ve}_{1} + {\ve}_{2}$. 

Also, one can formulate the $\Omega$-deformed theories on more general four-manifolds $X^{4}$, it suffices to have a $U(1)$, or $U(1) \times U(1)$ isometry (it should also be possible to extend this definition to the manifolds with $U(1)$-invariant conformal structure, but we shall not discuss this here). 
The idea is to use the $X^{4}$ bundle over ${\bf T}^{2}$ at the step ii.) of the procedure above, twisted by the elements of the symmetry group of $X^{4}$, which tend to identity as $r \to 0$. 

In technical terms, this procedure amounts to replacing the adjoint scalar ${\si}$ in the ${\CalN}=2$ four dimensional vector multiplet by the operator:
\begin{equation}
{\si} \mapsto {\si} +  {\ve}_1 {\nabla}_{{\varphi}_{1}} + 
{\ve}_{2} {\nabla}_{{\varphi}_{2}}
\label{eq:sicd}
\end{equation}
where ${\partial}_{{\varphi}_{1}}$ and ${\partial}_{{\varphi}_{2}}$ are the two vector fields on $X^{4}$ generating the $U(1) \times U(1)$ action. 

In this paper, as in \cite{Nekrasov:2009rc}, we shall be interested in a particular case of the 
$\Omega$-deformation, where ${\ve}_{2} = 0$. In this case the four dimensional ${\CalN}=2$ super-Poincare invariance is reduced to the two dimensional ${\CalN}=2$ super-Poincare. 
The parameter ${\ve}_{1} = {\hbar}$ is identified, in the spirit of \cite{Nekrasov:2009uh}, \cite{Nekrasov:2009ui} with the Planck constant of a quantum integrable system, obtained by the quantization of the Seiberg-Witten integrable system corresponding to $T_{4}$. 

One can think about this correspondence as a simplified version of the M-theory/string theory correspondence 
\cite{Witten:1995ex}, in a sense that the Planck constant of one theory is mapped to the geometric parameter of the other.

\subsection{Two dimensional flat connections and higher dimensional gauge theory}

The moduli space ${\CalM}_{g,n ; {\n}}$ of flat connections (or, which is the same, local systems)  with the gauge group $G$, on a genus $g$ Riemann surface $\Sigma$ with a finite number 
$n$ of punctures, with the prescribed conjugacy classes $\n$ of the monodromies around the punctures, is a frequent player in the studies of two, and three dimensional gauge theories, such as the Yang-Mills theory in two dimensions \cite{Atiyah:1982}, 
\cite{Witten:1992}, and the Chern-Simons theory in three dimensions \cite{Witten:1989}. In this context the case of a compact group $G$ is the most natural. 

In the attempts to describe the two \cite{Polyakov:1987}, \cite{Alekseev:1988ce}, \cite{Verlinde:1990}, \cite{Verlinde:1990h}  or three \cite{Witten:1988hc} dimensional quantum gravity using the formalism of vierbeins and spin connections the non-compact gauge groups, such as $SL(2, {\bR})$ or $SL(2, {\bC})$, become important. 

Despite some progress \cite{Chekhov:1999tn}, \cite{Chekhov:2000tw} along these lines the satisfactory
gauge theory construction of the quantum gravity theory
is still missing \cite{Witten:2007kt}.

Recently, the moduli spaces of flat connections ${\CalM}_{g,n; {{\n}}}$ on a Riemann surface $\Sigma$, with or without punctures, with the complex gauge groups, such as $SL(2, {\bC})$ or $PGL(2, {\bC})$ became ubiquitous in the study of the four dimensional ${\CalN}=2$ supersymmetric gauge theories, obtained by the compactification of the six dimensional $(0,2)$-superconformal theory of the $A_1$ type, on a Riemann surface $\Sigma$. In what follows we shall often use a shorter notation ${\CalM}^{loc}_{\Sigma}$ or ${\CalM}^{loc}$ for ${\CalM}_{g,n; {{\n}}}$. 

A simple way of seeing the role of ${\CalM}^{loc}_{\Sigma}$
is to compactify the theory on a circle ${\bf S}^{1}_{R_{a}}$, of radius $R_{a}$. On the one hand, the analysis of \cite{Seiberg:1996nz} shows that the effective description of the resulting theory is the three dimensional ${\CalN}=4$ supersymmetric sigma model on a manifold $M_{\rm tot}$ which is the total space of the Seiberg-Witten fibration over the moduli space of vacua of the four dimensional theory, in other words, the complex phase space of the Seiberg-Witten integrable system. 
On the other hand, remembering the six dimensional (0,2) origin of the theory in four dimensions, changing the order of compactification on $\Sigma$ and ${\bf S}^{1}_{R}$ we arrive at the picture where the five dimensional maximally supersymmetric Yang-Mills theory with the gauge group $SU(2)$ (for the general A,D,E type (0,2) theory one gets the A,D,E type Lie group as the gauge group), and the five dimensional coupling 
$$
{\bf g}_{5}^{2} = R_{a}
$$
(so that the Yang-Mills instantons, which are the solitons of the $4+1$ dimensional theory, could be identified with the Kaluza-Klein modes of the six dimensional theory), is further compactified with a partial twist on $\Sigma$. The partial twist makes two out of five adjoint scalars in the vector multiplet a one-form on $\Sigma$ valued in the adjoint bundle. 

In the limit of
vanishing size of $\Sigma$ we arrive at the three dimensional theory which is clearly the sigma model on the moduli space of the minimal energy configurations, which are the complex connections, ${\CalA} = A  + i {\phi}$, which are flat, ${\CalF} = d{\CalA} + {\CalA} \wedge {\CalA} = 0$, which are also D-flat, $D_{A}^{*}{\phi} = 0$, considered modulo
gauge transformations. The D-flatness condition and the compact gauge transformations together can be traded for the invariance under the complex gauge transformations. Thus, we end up with the sigma model on the moduli space 
${\CalM}_{\Sigma}^{loc}$ of complex flat connections on $\Sigma$, also known as the $G_{\bC}$-local systems.
The kinetic term of this sigma model is proportional to
$\frac{1}{{\bf g}_{5}^{2}} = \frac{1}{R_{a}}$, thus we are led to the conclusion, as in \cite{Seiberg:1996nz}, that the K\"ahler class of the metric on ${\CalM}$ as seen by the effective action is proportional to $\frac{1}{R_{a}}$.  The arguments of \cite{Seiberg:1996nz} involve the electric-magnetic duality, and this is the way to relate the two points of view we just reviewed.

\subsection{The hyperk\"ahler structure}

The two descriptions of the effective target space above are consistent. They just exhibit different complex structures on the same manifold. The target space of the ${\CalN}=4$ sigma model in three dimensions has to be a hyperk\"ahler manifold. For the theories we consider this manifold is the moduli space ${\CalM}_{H}$ of solutions of Hitchin's equations \cite{Hitchin:1987s}:
\begin{equation}
\begin{matrix}
& D_{z} {\phi}_{\bar z} = 0, \, D_{\bar z} {\phi}_{z} = 0, \\
& F_{z{\bar z}} + [ {\phi}_{z}, {\phi}_{\bar z} ]  = 0
\end{matrix}
\label{eq:hitchin}
\end{equation} 
which is indeed hyperk\"ahler (see \cite{Kapustin:2006pk}
for the detailed review of its properties). 
In the complex structure (which is conventionally denoted by $I$) where the components $A_{\bar z}, {\phi}_{z}$ of the gauge field and the twisted scalar are holomorphic, the space ${\CalM}_{H}$ has the structure of the algebraic integrable system \cite{Hitchin:1987i}, with the base being the space of holomorphic differentials
of degrees $d_{1}, d_{2}, \ldots, d_{r}$, for $r = rk(G)$, 
$P_{i}({\phi}_{z}) \in H^{0}({\Sigma}, K_{\Sigma}^{d_{i}})$
for the degree $d_{i}$ invariant polynomials on $Lie(G)$, and the fiber being a (complement to a divisor in a) Jacobian of the spectral curve ${\CalC} \subset T^{*}{\Sigma}$, defined by the equation (for $G = SU(N)$, for other Lie groups see the review \cite{Donagi:1995s}) 
$$
{\rm Det}({\phi}_{z} - {\lam}) = 0 
$$
In the complex structure $J$, where the holomorphic coordinates are the components 
$$
{\CalA}_{z} = A_{z} + i {\phi}_{z}, 
\qquad {\CalA}_{\bar z} = A_{\bar z} + i {\phi}_{\bar z}, 
$$
$M_{H}$ is identified (up to the usual stability issues) with the moduli space of the complex $G_{\bC}$-flat connections. 
Finally, $K = IJ$, and the $K$-holomorphic coordinates are
$$
A_{z} + {\phi}_{z}, \qquad A_{\bar z} - {\phi}_{\bar z}
$$
The complex structure $J$ is natural, if one thinks of the three dimensional theory as coming from the compactification of the six dimensional ${\CalN}=1$ gauge theory on a three manifold ${\Sigma} \times {\bf S}^{1}_{r^{\prime}}$, in the limit $r^{\prime} \to \infty$.  

To say that ${\CalM}_{H}$ is hyperk\"ahler means that there exists the whole two-sphere of complex structures, 
$$
{\CalI} = a I + b J + c K, \quad {\CalI}^2 = - 1, 
$$  
for any $(a,b,c)$, s.t. $a^{2} + b^{2} + c^{2} = 1$,
and the two-sphere of the corresponding K\"ahler forms, 
\begin{equation}
\label{eq:kahlrf}
{\om}_{{\CalI}} = a {\om}_{I} + b {\om}_{J} + c {\om}_{K}
\end{equation}
where
\begin{equation}
{\om}_I = \int_{\Sigma} {\rm tr}\, \left( {\de} A \wedge {\de} A - {\de} {\phi} \wedge {\de} {\phi} \right) 
\label{eq:omi}
\end{equation}
\begin{equation}
{\om}_J = \int_{\Sigma} {\rm tr}\, \left( {\de} A \wedge \ast{\de} \phi  \right) 
\label{eq:omj}
\end{equation}
\begin{equation}
{\om}_K = \int_{\Sigma} {\rm tr}\, \left( {\de} A \wedge {\de} \phi  \right) 
\label{eq:omk}
\end{equation}
For the compact $\Sigma$ the form ${\om}_{I}$ realizes a nontrivial cohomology class of ${\CalM}_{H}$, while ${\om}_{J}$ and ${\om}_{K}$ are cohomologically trivial. 
We shall normalize the forms (\ref{eq:kahlrf}) in such a way that ${\om}_{I}$ realizes the integral cohomology class, the restriction of ${\om}_{I}$ onto the subvariety ${\rm Bun}_{G}$
where ${\phi} = 0$ is, up to the $(2\pi i)$ multiple, the curvature of the canonical Hermitian connection on the determinant line bundle ${\sl L}$ over ${\rm Bun}_{G}$:
\begin{equation}
[ {\om}_{I} ] \biggr\vert_{{\rm Bun}_{G}}   = c_{1}\left( {\sl L} \right)
\label{eq:fstch}
\end{equation}
If the Riemann surface $\Sigma$ has $n > 0$ punctures, then all three symplectic forms ${\om}_{I,J,K}$ 
 on the moduli space of the solutions to Hitchin's equations with the sources are, in general, 
 cohomologically non-trivial. 
 
\subsection{From ${\CalN}=2$ 4d gauge theory to 2d ${\CalN}=(4,4)$ sigma model}

We can further compactify the theory on a circle ${\bf S}^{1}_{R_{b}}$. On the one hand, this gives a two dimensional
${\CalN}=(4,4)$ sigma model with the same target space $M_{H}$, whose K\"ahler class is proportional to
$R_{b}/R_{a}$. More generally, if we start with the $(0,2)$
theory and compactify it on $E_{\rho} \times \Sigma$, where
$E_{\rho}$ is the elliptic curve with the complex modulus $\rho$, we would get the two dimensional sigma model on $M_{H}$ with the complexified K\"ahler class
\begin{equation}
[{\varpi}] = {\rho} [{\om}_{I}]
\label{eq:rhok}
\end{equation}
The two dimensional ${\CalN}=2$ sigma model can be topologically twisted to define an $A$ model, which depends on the symplectic structure of the target space, or to define a $B$ model, which depends on the complex structure of the target space. 
The ${\CalN}=(2,2)$ theory can be twisted in an asymmetric manner, so that the left- and the right- chirality worldsheet one-form fermions would transform into the ${\bar\partial}$-derivatives of the worldsheet bosons
which are holomorphic in the target space in two different complex structures $(I_{+}, I_{-})$ (there are also the generalizations involving the generalized complex structures but we shall not need them, see the discussion and the references in \cite{Kapustin:2006pk}). 

\subsection{$\Omega$-deformation as the boundary condition}

In topology, the $G$-equivariant cohomology of a space $Y$ with the free $G$-action is the ordinary cohomology of the quotient space $Y/G$. As a module over $H^{*}(BG)$ (which is a polynomial ring) this is a pure torsion, and so, upon localization becomes trivial. 

Similarly, the $\Omega$-deformation of the gauge theory living on a spacetime $X^4$ with the free acting $U(1)\times U(1)$ isometry is can be undone by a field redefinition, and a redefinition of the couplings, as explained, for $X^4 = T^2 \times B^2$, in \cite{Nekrasov:2010ka}. On the other hand, as we reviewed in the previous section, the four dimensional gauge theory compactified on $T^2$, reduces, at low energy, to the sigma model on the total space of the Seiberg-Witten fibration. The relation (\ref{eq:rhok}) gets modified, as the effective K\"ahler parameter $\rho$ and the effective complex structure parameter $\tau$ (the asymptotic gauge coupling) get mixed with the parameters ${\ve}_{1}, {\ve}_{2}$ of the $\Omega$-deformation. 

If $U(1) \times U(1)$ acts on the four-manifold $X^4$ with the fixed points, then the transformation mapping the 
gauge theory to the sigma model is valid approximately, outside the fixed locus. We can view the worldsheet $B^2 = X^4/T^2$ of the sigma model as a two-manifold with corners, perhaps non-compact, and replace the effects of the $\Omega$-deformation, which cannot be removed near the boundary ${\p}B^2$, by some boundary conditions, i.e. the branes in the sigma model.

It is argued in \cite{Nekrasov:2010ka} that at the smooth connected component of the boundary ${\p}B^2$ the corresponding brane can be interpreted as a $(A, B, A)$ brane, which, depending on the duality frame, is either the so-called canonical coisotropic brane (the cc-brane ${\CalB}^{cc}$, for short), or a particular complex (in the complex structure $J$) Lagrangian (with respect to ${\om}_{I}$ and ${\om}_K$) brane, known as the brane of opers ${\CalB}_{{\CalO}_{\tau}}$, in the case of the $SU(2)$ Hitchin moduli space ${\CalM}_{H}$. This brane depends on the complex structure ${\tau}$ of the Riemann surface 
$\Sigma$, while the topological field theory, in which it is BRST invariant, does not. 

The components at infinity ${\p}B^2_{\infty} = {\overline B^2} \backslash B^2$ also come with some boundary conditions, both in the gauge theory, and in the effective sigma model. It was argued in \cite{Nekrasov:2010ka} that the corresponding branes ${\CalB}^{\infty}_{\gamma}$
also correspond to the middle-dimensional complex Lagrangians ${\CalL}_{\gamma}$ of the $(A,B,A)$-type, which, unlike ${\CalO}_{\tau}$, are defined in topological terms (i.e. these do not depend on the complex structure of $\Sigma$), yet may depend on some combinatorial data $\ga$. 

From now on we shall be discussing the $A_{1}$ type theories. 
To describe the branes ${\CalB}^{cc}, {\CalB}_{{\CalO}_{\tau}}, {\CalB}^{\infty}_{\gamma}$ in more detail we need to discuss the geometry of the moduli space of flat connections. 

\subsection{Two two dimensional theories}

Suppose we study the four dimensional theory on the manifold $X^4 = C^2 \times {\bf S}^1 \times {\bf R}^1$, where $C^2$ is topologically a disk, with the cigar metric:
\begin{equation}
\label{eq:cigar}
ds^2_{C^2} = dr^2 + f(r) d{\varphi}_{1}^{2}
\end{equation}
Where $f(r) \sim r^2$ for $r \to 0$ and $f(r) \sim R^2$, for $ r \to \infty$, for some constant $R$.  
Let ${\varphi}_{2}$ be the angular coordinate on the second ${\bf S}^{1}$. The base $B^2$ is, in this case, the half-plane ${\bf R}_{+} \times {\bf R}$. 
Suppose we turn on the $\Omega$-deformation corresponding to the isometry rotating $C^2$, i.e. the isometry generated by ${\p}/{\p}{\varphi}_{1}$. 

On the one hand, we can relate this theory to the sigma model with the worldsheet $B^2$, as in \cite{Nekrasov:2010ka}. 
The sigma model brane corresponding to the boundary $r=0$ is ${\CalB}^{cc}$ or its T-dual ${\CalB}_{{\CalO}_{\tau}}$.
The other "boundary", at $r \to \infty$, leads to some asymptotic boundary condition, which we view as the brane ${\CalB}^{\infty}$, or ${\CalB}^{\infty}_{\gamma}$ if we want to specify the type $\gamma$ of the boundary conditions.  We shall call this viewpoint the theory $T_{rt}$

On the other hand, we can view this theory as a two dimensional ${\CalN}=(2,2)$
gauge theory with the worldsheet ${\bf S}^{1} \times {\bf R}^{1}$, with an infinite number of fields, as in \cite{Nekrasov:2009rc}.  
The $\Omega$-deformation corresponds, in this two dimensional language, to turning on the twisted mass
$\ve$, corresponding to the global symmetry $U(1)$, which is the rotation of $C^2$. We shall call this two dimensional theory the theory $T_{2t}$. 

The two dimensional theory has at low energies (in the sense of the theory on ${\bf S}^{1} \times {\bf R}^{1}$) a description of the sigma model on the complexified Cartan subalgebra ${\bf t}_{\bC}$ of the gauge group, with the effective twisted superpotential ${\tilde\CalW}^{\rm eff} ({\si}; {\tau};  m, {\ve})$. Here ${\si}$ denotes the flat coordinates on 
${\bf t}_{\bC}$, ${\tau}$ denotes the four dimensional complexified gauge couplings, which are identified, for the  $A_1$ theories, with the complex moduli of $\Sigma$ in a suitable parametrization, $m$ denotes the masses of the matter multiplets in four dimensions, and finally ${\ve}$ is the parameter of the $\Omega$-background which is viewed as the two dimensional twisted mass.  This superpotential can be split as a sum of two contributions: a contribution of the fixed point in $C^2$
and a contribution of the boundary at infinity:
\begin{multline}
\label{eq:fixp}
{\tilde\CalW}^{\rm eff} ({\si}, {\tau},  m, {\ve}, {\ga}) = \\
\quad {\ve} \left( {\CalW}_{{\CalO}_{\tau}} ({\si}/{\ve}, m/{\ve}) - {\CalW}_{\infty} ({\si}/{\ve}, m/{\ve}, {\ga})  \right) \\
\end{multline}
The contribution ${\CalW}_{{\CalO}_{\tau}} ({\si}/{\ve}; m/{\ve})$ of the fixed point is computed from the asymptotics of the ${\CalZ}$-function as follows \cite{Nekrasov:2009rc}:
\begin{equation}
\label{eq:wfromz}
\begin{split}
{\CalW}_{{\CalO}_{\tau}}( {\al}, {\n} ) = \qquad\qquad\qquad\qquad\qquad\qquad \\
\quad \frac{1}{\ve} {\rm Lim}_{{\ve}_{2} \to 0}\ {\ve}_{2} {\rm log}{\CalZ} ( {\al}{\ve}, {\tau}, {\n}{\ve} ; {\ve}, {\ve}_{2}) \\
\end{split}
\end{equation}
The left hand side is independent of $\ve$, as follows from the asymptotic conformal invariance of the gauge theory under consideration. 

The contribution of the region at infinity ${\CalW}_{\infty}({\al}; {\n}, {\ga})$ is independent of
$\tau$. This is demonstrated by observing \cite{Nekrasov:2002qd} that the gauge theory subject to the $\Omega$-deformation is extremely weakly coupled far away from the fixed points of the action of the isometry involved in the $\Omega$-background. So we do not expect anything interesting to come from the bulk, however, if we cut the cigar $C^2$ at some finite value of $r$ to have an infrared regulator, then we may expect some effective three dimensional supersymmetric gauge theory to live on the product of the circle at infinity ${\bf S}^{1}_{\infty}$ and the cylinder ${\bf S}^{1}\times {\bf R}^{1}$, in analogy with the analysis of \cite{Gaiotto:2008bc}, \cite{Gaiotto:2008bd}. Compared to the situation studied there we have half the supersymmetry, so the three dimensional theory has only four supercharges, and upon compactification on ${\bf S}^{1}_{\infty}$ generates a twisted superpotential, 
of the form studied in \cite{Nekrasov:2009uh},\cite{Nekrasov:2009zz}
\begin{equation}
{\CalW}_{\infty} ({\al}; {\nu}, {\ga}) \sim \sum_{\ell}
Li_{2}( e^{{\ell}({\al}, {\n})} )
\label{eq:dilw}
\end{equation}
where the sum is over the charged matter fields, and ${\ell}$ is the linear function of the gauge multiplet scalar vevs and the masses. The degrees of freedom living at ${\bf S}^1_{\infty}$ depend on some combinatorial data $\ga$ which we shall make more explicit later, but 
we shall not attempt to identify the boundary theory and the corresponding superpotential more precisely. Instead, we focus on ${\CalW}_{{\CalO}_{\tau}}$.
  
\subsection{Twisted superpotential as the generating function}

Our main conjecture is as follows:

{\em The effective twisted superpotential of the theory 
on ${\bf S}^{1} \times {\bf R}^{1}$ obtained by localizing at the fixed point in $C^2$ is essentially the difference of the generating functions of the Lagrangian subvarieties ${\CalO}_{\tau}$ and ${\CalL}_{\gamma}$
in ${\CalM}^{loc}$, defined with respect to the appropriate Darboux coordinate system on ${\CalM}^{loc}$. The supersymmetric vacua of the theory $T_{2t}$ correspond to the intersection points $v \in {\CalO}_{\tau} \cap {\CalL}_{\ga}$, which are also the vacua of the theory $T_{rt}$ subject to the appropriate D-brane boundary conditions.}

This statement can be viewed as the improvement on the result of A.~Beilinson and V.~Drinfeld. They show in 
 \cite{BeDrH} that upon the holomorphic quantization of the Hitchin system for the group $G$ the spectrum (in the sense of commutative algebra) of twisted differential operators, e.g. the abstract quantum commuting Hamiltonians, identifies canonically with ${\CalO}_{\tau}$ for the dual group $^{L}G$. In this paper we will concentrate on 
$$
G=PGL(2,C), \,  ^{L}G=SL(2,C)\ .
$$   
Our point is that
in the Darboux coordinate system $({\al}, {\be})$ the generating function ${\CalW}_{{\CalO}_{\tau}}({\al}, {{\n}})$ of the variety of opers, 
\begin{equation}
{\be} = \frac{{\p}{\CalW}_{{\CalO}_{\tau}} ({\al}, {{\n}})}{{\p}{\al}}
\label{eq:opers}
\end{equation} is essentially the Yang-Yang function of the quantum Hitchin system. More precisely, the Yang-Yang function ${\CalY}({\al}, {{\n}}, {\tau}, {\ga})$ of the quantum Hitchin system depends on the complex structure
${\tau}$ of $\Sigma$ (as does the Hitchin's integrable system) and on the choice of the "real slice", which defines the space of states ${\CalH}_{\ga}$. Our claim is:
\begin{equation}
{\CalY}({\al}, {{\n}},  {\tau}; {\ga}) = {\CalW}_{{\CalO}_{\tau}}({\al}, {{\n}}) - {\CalW}_{\ga} ({\al}, {{\n}})
\label{eq:yyh}
\end{equation} 
i.e. up to a ${\tau}$-independent piece the twisted superpotential ${\CalW}_{{\CalO}_{\tau}}({\al}, {{\n}})$ computed by the four dimensional instanton calculus (\ref{eq:wfromz}) is the Yang-Yang function. 
Indeed, the coordinates $({\al}, {\be})$ are defined up to $2{\pi}i{\bZ}$, so the {\em Bethe} equation
\begin{equation}
\frac{{\p}{\CalY}({\al}, {{\n}}, {\tau}, {\ga})}{{\p}{\al}_k} = 2{\pi} i n_k , \, n_k \in {\bZ}
\label{eq:int}
\end{equation}
determines the spectrum 
\begin{equation}
E_k ({\vec n}) = \frac{{\p}{\CalY}({\al}, {{\n}}, {\tau}, {\ga})}{{\p}{\tau}_k} 
\label{eq:enk}
\end{equation}
of the quantum Hitchin system. Let us clarify the meaning of (\ref{eq:enk}). The classical Hamiltonians of the $A_1$ type Hitchin system are the quadratic differentials with the second order poles at the punctures, with the prescribed leading singularity. More precisely, given a basis 
${\mu}_{(k){\bar z}}^{z}$, $k = 1, \ldots , 3g-3+n$, of the Beltrami differentials which correspond to the variations of the complex structure of the punctured $\Sigma$, 
\begin{equation}
{\mu}_{(k){\bar z}}^{z} \leftrightarrow \frac{\p}{{\p}{\tau}_{k}}
\label{eq:belt}
\end{equation} 
we compute:
\begin{equation}
\label{eq:hamh}
H_{k} = \int_{\Sigma} {\mu}_{(k){\bar z}}^{z} {\rm Tr}{\phi}_{z}^{2}
\end{equation}
Upon the deformation quantization $H_k$ become the elements of the noncommutative ring (one has to talk about the sheaf of $D$-modules to actually see the noncommutative algebras, since the globally defined objects form a commutative subalgebra, \cite{BeDrH}), whose spectrum we wish to determine. One has to specify the space of states on which we represent both the noncommutative algebra and its commutative subalgebra of the quantum integrals of motion. This is done (indirectly) by picking up a Lagrangian submanifold ${\CalL}_{\ga}$. At the moment it is hard to make this construction explicit, since it involves the T-duality along the fibers of the Hitchin fibration. See \cite{Gukov:2008bq},\cite{Nekrasov:2010ka}
for more details. 
The formula (\ref{eq:enk}) actually makes sense even without specifying the space of states. It reflects the canonical identification \cite{BeDrH} of the spectrum of the commutative algebra of the quantized Hitchin's Hamiltonians (the twisted differential operators on $Bun_G$) with the variety of opers. Indeed, ${\CalO}_{\tau}$ is a Lagrangian submanifold in ${\CalM}^{loc}$. As we vary the complex structure ${\tau}$ infinitesimally, the corresponding variation of the Lagrangian submanifold
is described by a closed one-form defined on ${\CalO}_{\tau}$, i.e. a holomorphic function, since the variety of opers is simply-connected. This function can be computed locally using the Hamilton-Jacobi equation which gives
(\ref{eq:enk}) in the "static gauge" where to view the deformation of
${\CalO}_{\tau}$ while keeping ${\al}$, the "Lagrangian half of the monodromy data", fixed.

Let us conclude with a few comments. 
In the context of the quantum integrable systems the Yang-Yang function is defined as a potential for Bethe equations, which is unique up to a constant which could be function of the parameters of the system, such as the complex structure parameters $\tau$ in our case. This ambiguity can be partly fixed by requiring that the derivatives of the Yang-Yang function with respect to the parameters ${\tau}_{k}$ correspond to the suitably normalized operators ${\Phi}_{k}$, forming  a basis in the space of quantum integrals of motion.

The eigenvalue $E_k ({\vec n})$ in (\ref{eq:enk}) is calculated on the solutions of (\ref{eq:int}) and depends on the discrete parameters ${\vec n}$. Obviously, the ambiguity we referred to above does not affect the differences $E_k ({\vec n}) - E_k ({\vec n}')$ of levels.

\bigskip
We would like to stress that both Eqs. (\ref{eq:int}), (\ref{eq:enk}) make sense for any choice of the Darboux coordinates on ${\CalM}^{loc}$, as they express in coordinates two geometric facts: i) The eigenstates of the quantum system correspond to the intersection points $v$ of two Lagrangian submanifolds in ${\CalM}^{loc}$; ii) the eigenvalues of the quantum Hamiltonians are the functions on the variety of opers which generate its deformations corresponding to the variations of the complex structure of $\Sigma$, evaluated at the intersection points $v$. 
The way we presented these equations uses the understanding of the Bethe/Gauge correspondence \cite{Nekrasov:2009uh}, \cite{Nekrasov:2009zz}, \cite{Nekrasov:2009ui} developed in \cite{Nekrasov:2009rc}
for the case of the two dimensional $\Omega$-background. Another,
closely connected relation, was developed in \cite{Alday:2009aq}
 where it has been demonstrated that in the 
general $\Omega$-background the ${\CalZ}$-function gives the conformal block of some
two dimensional conformal field theory, which were identified for a large class of gauge theories. The two dimensional $\Omega$-background
maps to the classical limit of the \cite{Alday:2009aq} correspondence on the conformal field theory side, which is obvious from the definitions (\ref{eq:wfromz}), (\ref{eq:yyh}) and
expressions  (\ref{eq:int}), (\ref{eq:enk}). Thus it should be possible to see (\ref{eq:enk}), which states that $E_k$ can be
expressed in terms of some potential, in any Darboux coordinate system, as
follows from the interpretation given in \cite{Nekrasov:2010ka}, purely in conformal field theory without any use of  \cite{Alday:2009aq}, which was indeed done in \cite{Teschner:2010qh} for genus
zero with punctures. Stated this way, purely in the language of conformal field theory, the relation
(\ref{eq:enk}) reminds another relation in classical Liouville theory
(for special values of $\nu$) �known from \cite{Takhtajan:1988} with ${\CalY}$ being replaced by the
{\it holomorphic part of the classical Liouville action}. These
relations with \cite{Teschner:2010qh}, as well as with other CFT papers mentioned above, must have profound
consequences.

In this sense the equations (\ref{eq:int}), 
(\ref{eq:enk}) do not determine the Yang-Yang function ${\CalY}({\al}, {\tau}, {\n}, {\ga})$. 
However, once both ${\al}$ and ${\be}$ coordinates are fixed, the Yang-Yang function is determined by the Lagrangian submanifolds ${\CalO}_{\tau}$
and ${\CalL}_{\ga}$, and the claim that the $\tau$-dependent piece coincides with the localized four dimensional gauge 
theory twisted superpotential (\ref{eq:wfromz}) becomes quite nontrivial. It would not be true if instead of our $({\al}, {\be})$ coordinates one took, e.g.,
Fock-Goncharov coordinates \cite{FoGo}, which are used in \cite{Gaiotto:2009hg} with much success. 

\section{Geometry of the moduli space of flat connections}

Recall that we use the notations ${\CalM}_{g,n ; {{\n}}}$, ${\CalM}_{\Sigma}^{loc}$ or ${\CalM}^{loc}$ interchangeably. 

\subsection{Moduli of flat connections: explicit description}
 
The moduli space ${\CalM}_{g,n ; {{\n}}}$ of flat connections is the space of equivalence classes of the homomorphisms of the fundamental group of the $n$-punctured genus $g$ surface to the gauge group $G$, with the condition that the conjugacy class $[ {\gm}_{k}] $ of the image of the simple loop ${\gm}_{k}$ surrounding the $k$'th puncture lands in a particular conjugacy class in $G$, which we label by ${\n}_{k}$. 
In case $G = SL(2, {\bC})$ we view ${\n}_{k}$ as a complex number modulo the action of the affine Weyl group which flips the sign of ${\n}_{k}$ and shifts it by an integer. The invariant is the trace $m_{k} \in {\bC}$ of the monodromy ${\gm}_{k}$ around the puncture $z_{k}$:
\begin{equation}
\label{eq:mtr}
m_{k} = 2\, {\rm cos} (2 \pi {\n}_{k}) = {\tr} ( {\gm}_{k} )
\end{equation}
To be more precise we shall study the moduli space of flat $G$ connections on a surface $\Sigma$ with $n$ small disks removed. This moduli space can be identified, very simply, with the
space of $(2g+n)$-tuples of the elements of $G$, $({\ba}_{1}, {\bb}_{1}, \ldots, {\ba}_{g}, {\bb}_{g}; {\gm}_{1}, \ldots, {\gm}_{n})$, obeying:
\begin{equation}
\label{eq:mmom}
 {\gm}_{n}{\gm}_{n-1} \ldots {\gm}_{2}{\gm}_{1} \prod_{l=1}^{g} {\ba}_{l}{\bb}_{l}{\ba}_{l}^{-1}{\bb}_{l}^{-1}  = 1
\end{equation}
considered up to a simultaneous conjugation:
\begin{equation}
\label{eq:fltcn}
( {\ba}_{1}, {\bb}_{1}, \ldots , {\ba}_{g}, {\bb}_{g}; {\gm}_{1}, \ldots , {\gm}_{n}) \sim
\end{equation}
$$
\quad 
(h^{-1}{\ba}_{1}h, h^{-1}{\bb}_{1}h, \ldots , h^{-1}{\ba}_{g}h, h^{-1}{\bb}_{g}h; 
$$
$$ h^{-1}{\gm}_{1}h, \ldots , h^{-1}{\gm}_{n} h), 
$$
for any $h \in G$, and with the fixed conjugacy classes, which for $G = SL(2, {\bC})$ reads as ${\tr} ( {\gm}_{k} ) = 2\, {\rm cos} (2{\pi}{\n}_{k} )$. 

This identification depends upon the choice of the generators of the fundamental group of the complement
${\Sigma} \backslash \{ z_{1}, \ldots , z_{n} \}$, as in the Fig. \ref{fig:genfund}:
\begin{figure}
\includegraphics[width=15pc]{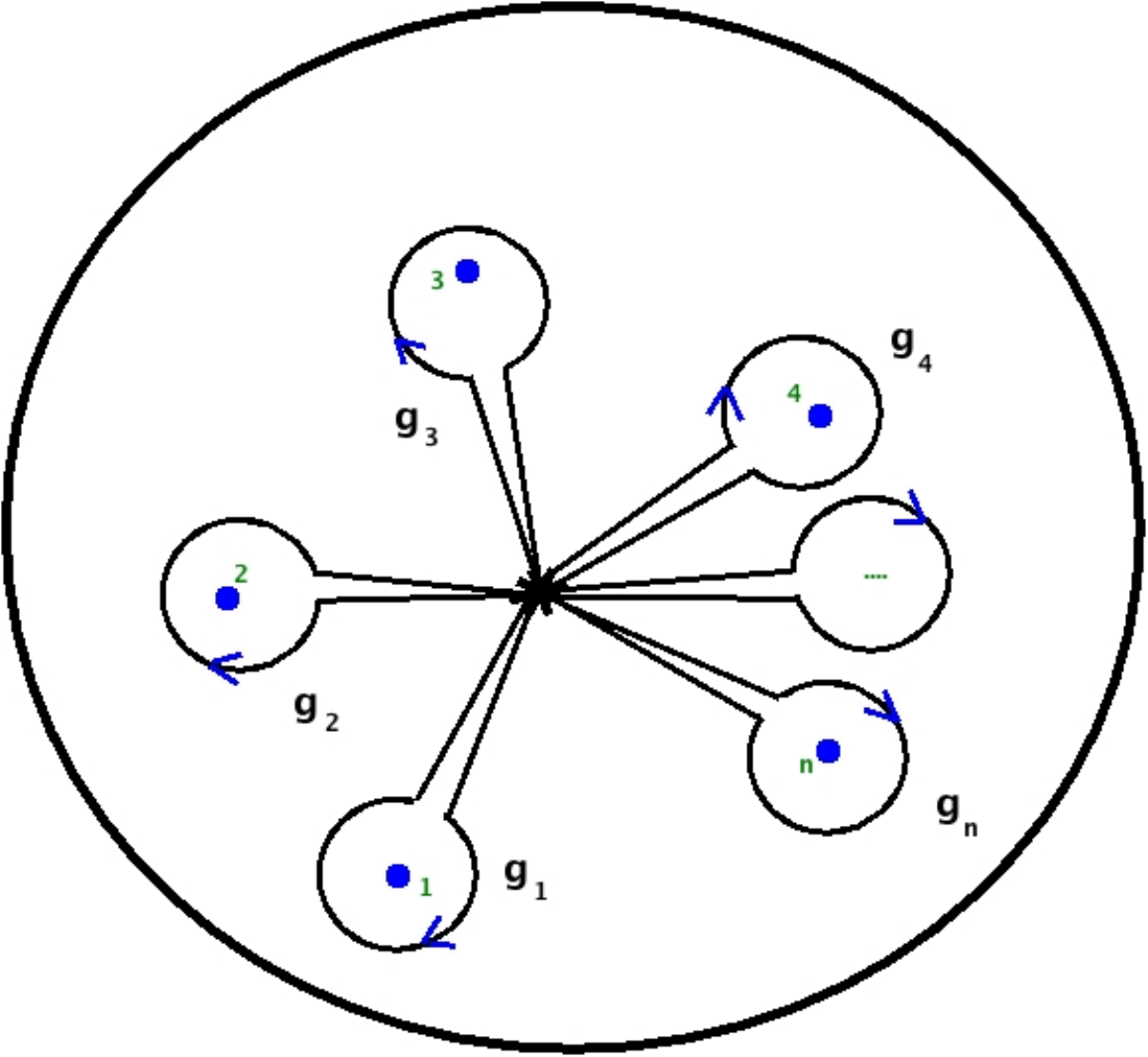}
\caption{Generators of  ${\pi}_{1} \left( {\bf S}^{2} \backslash \{ z_{1}, \ldots , z_{n} \} \right)$}
\label{fig:genfund}
\end{figure}
From now we assume $({\n}_{1}, \ldots, {\n}_{n})$ which allows us to think of the space 
 of functions on ${\CalM}_{g,n; {\n}}$
as of the polynomial ring generated 
by the traces of monodromies around the noncontractible loops. This could be viewed as a definition of the ${\CalM}_{g,n; {\n}}$ as of an affine variery. 

Alternatively, one could choose a subset of stable points. This choice depends, in turn, on the $n$-tuple of real numbers
$r_{1}, \ldots ,  r_{n}$. To implement this procedure properly one needs to use the full set of
Hitchin's equations \cite{Hitchin:1987s} with the delta function sources (see, e.g. \cite{Gorsky:1994dj}, \cite{Nekrasov:1995nq}, \cite{Donagi:1995am}, \cite{Kapustin:1998pb}, \cite{Markman:1994}, \cite{Gorsky:1999rb}). 
We shall not discuss this any further.

\subsection{The symplectic form}

The moduli space of flat connections on  a compact Riemann surface is a symplectic manifold \cite{Atiyah:1982}, with the symplectic form written on the space of all connections as:
\begin{equation}
\Omega = \int_{\Sigma} {\rm Tr} \, {\de}A \wedge {\de}A
\label{eq:flcnsf}
\end{equation}
When one studies the moduli of flat connections on a surface with boundaries, one may use the Poisson description. 
Also, in the finite dimensional description of the moduli space, e.g. as in (\ref{eq:mmom}) one can realize the Poisson structure on the moduli space as descending from that on the space of graph connections, \cite{Fock:1992xy},\cite{Fock:1993ms}, \cite{Fock:1997ai}.
Alternatively, one may use the formalism of 
\cite{FoGo} to describe the symplectic form and some set of Darboux coordinates (which are different from the coordinates $({\al}, {\be})$ which we introduce below!) associated with the triangulations of $\Sigma$ with marked points (this formalism works if there is at least one puncture). 
See also \cite{Goldman}, \cite{Goldman:1986if}, \cite{Turaev}, 
\cite{Tyurin}.

Using any of the formalisms above, or even the basic formula (\ref{eq:flcnsf}) which defines 
the Poisson
structure on the space of all connections one deduces  that the Poisson bracket of the Wilson loops in the fundamental representation is given by the "skein-relations" \cite{Goldman:1986if}, \cite{Turaev}:
\begin{equation}
\label{eq:psk}
\left\{ {\rm tr} P{\exp}\oint_{{\ga}_{1}} {\CalA}, \  {\rm tr} P{\exp}\oint_{{\ga}_{2}} {\CalA} \right\} = \qquad\qquad\qquad
\end{equation}
$$
{1\over 2}\sum_{x\in {\ga}_{1} \cap {\ga}_{2}}\, \left(  {\tr} P{\exp}\oint_{{\ga}^{+}_{x,1,2}}{\CalA} - {\rm tr} P{\exp}\oint_{{\ga}^{-}_{x,1,2}}{\CalA} \right) 
$$
where the loops ${\ga}_{x,1,2}^{\pm}$ are obtained by removing a small neighborhood of the intersection point $x$ and replacing it by two arcs making a single loop, as in the Fig.\ref{fig:skein}:
\begin{figure}
\includegraphics[width=15pc]{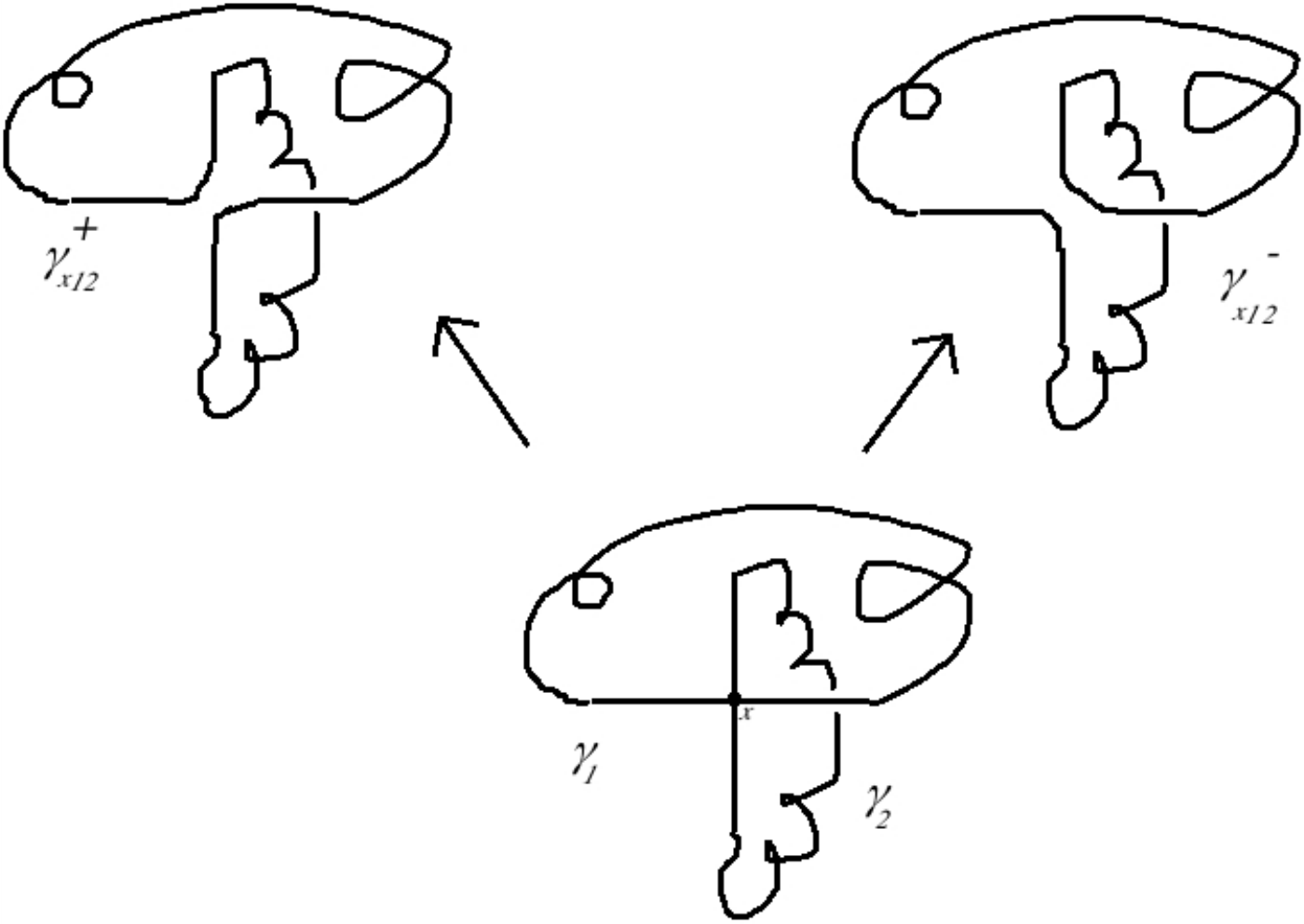}
\caption{The intersecting loops ${\ga}_{1}$, ${\ga}_{2}$ and the simple loops ${\ga}_{x,1,2}^{\pm}$}
\label{fig:skein}
\end{figure}

\subsubsection{The case of a sphere with four punctures}
\begin{figure}
\includegraphics[height=6pc]{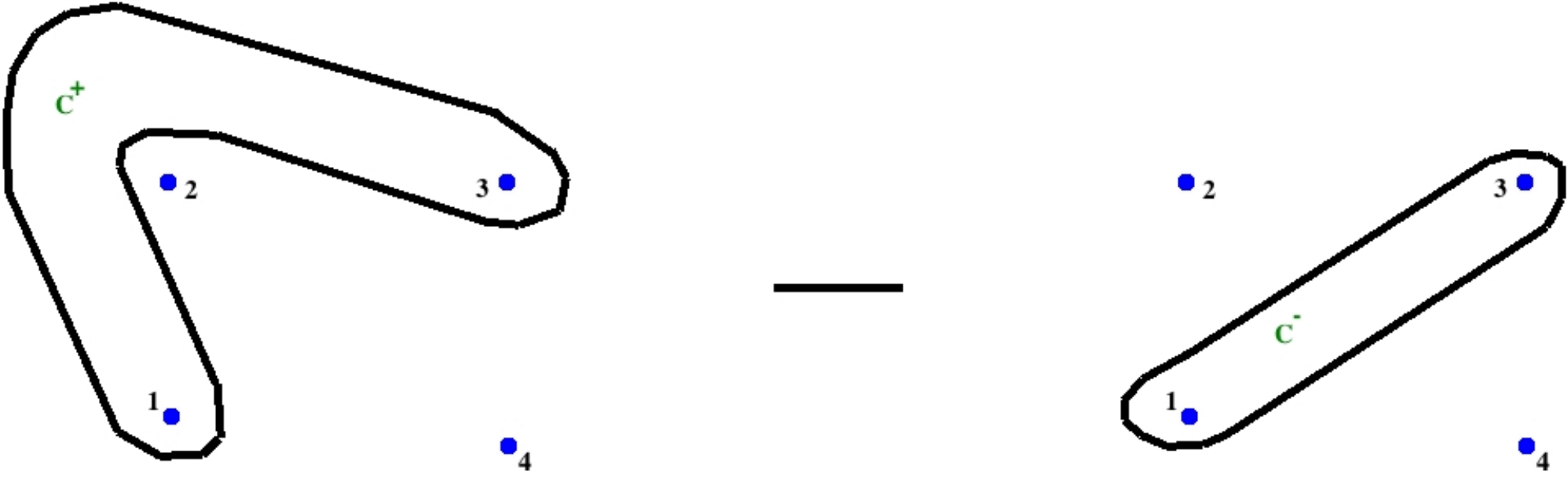}
\caption{The $C^{+}$ and $C^{-}$ loops}
\label{fig:PBg0}
\end{figure}

Let us study first the case of $g=0, n=4$ in some detail. Let ${{\n}} = \left( {\n}_{1}, {\n}_{2}, {\n}_{3}, {\n}_{4} \right)$.  The moduli space ${\CalM}_{0,4;{\n}}$ has the complex dimension two:
$$
{\CalM}_{0,4;{\n}} = \qquad\qquad\qquad\qquad\qquad\qquad\qquad\qquad
$$
\begin{equation}
\label{eq:modsp}
\left\{ \left( {\gm}_{1}, {\gm}_{2}, {\gm}_{3} \right) \in G^{\times 3} \ \biggr\vert  
\begin{matrix}& {\tr}({\gm}_{i}) = {m}_{i}, \  i = 1,2,3  \\
& {\tr} ({\gm}_{3}{\gm}_{2} {\gm}_{1}) = {m}_{4} \\
\end{matrix}
\right\} \, \biggr/ \, G 
\end{equation}
where $G$ acts by simultaneous conjugation $\left( {\gm}_{1}, {\gm}_{2}, {\gm}_{3} \right) \mapsto \left( h^{-1}{\gm}_{1}h, h^{-1}{\gm}_{2}h, h^{-1}{\gm}_{3}h \right)$.
The generators of the coordinate ring of ${\CalM}_{0,4;{\n}}$
can be taken to be:
\begin{equation}
\label{eq:genfr}
\begin{split}
A = m_{12} = {\tr}({\gm}_{1}{\gm}_{2}), \\ 
B = m_{23} = {\tr}({\gm}_{2}{\gm}_{3}), \\
C = m_{13} = {\tr} ( {\gm}_{1}{\gm}_{3} ) \\
\end{split}
\end{equation}
with one polynomial relation (which is easy to verify)
\begin{equation}
\label{eq:wof}
\begin{split}
 W_{0,4}(A,B,C) = 0, \qquad\qquad\qquad\qquad \\
 W_{0,4} = 
ABC + A^{2} + B^{2} + C^{2} - 4   \\
  - A (m_{3}m_{4} + m_{1} m_{2})  \\
  - B (m_{1}m_{4} + m_{2}m_{3} ) \\
  - C (m_{1}m_{3} + m_{2}m_{4} ) \\
  +  m_{1}^{2} + m_{2}^{2} + m_{3}^{2} + m_{4}^{2} + m_{1}m_{2}m_{3}m_{4} \\
\end{split}
\end{equation}
The application of the rule (\ref{eq:psk}) to the loops drawn around the pairs of points $z_{1}, z_{2}$ and $z_{2}, z_{3}$, respectively, 
as in the Fig.\ref{fig:ABloops}:
\begin{figure}
\includegraphics[width=12pc]{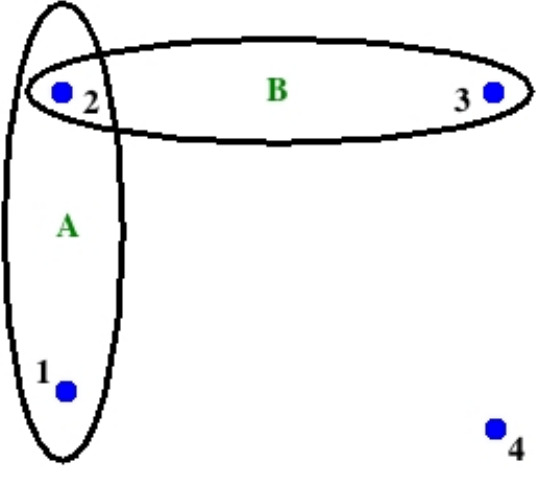}
\caption{The $A$ and $B$ loops}
\label{fig:ABloops}
\end{figure}
gives rise to the difference of two loops, 
$$
\left\{ A, B \right\} = C^{+} - C^{-}
$$
drawn on Fig. \ref{fig:PBg0}, 
which can be expressed via $A,B$, and $C$ as follows:
\begin{equation}
\label{eq:cpm}
C^{+} = {\tr}( {\gm}_{2}^{-1}{\gm}_{1}{\gm}_{2}{\gm}_{3} ) = 
\end{equation}
$$ 
\qquad - C - AB + m_{1}m_{3}+m_{2}m_{4} , \ 
$$
$$
C^{-}  = C
$$
and also as the derivative of $W_{0,4}$: 
\begin{equation}
\left\{ A, B \right\} = \frac{{\p}W_{0,4}}{{\p}C}
\label{eq:pbgz}
\end{equation}
In arriving at (\ref{eq:cpm}) we used the following identity for the $SL(2)$ matrices:
\begin{equation}
\label{eq:idsl}
{\tr}(g) {\tr}(h) = {\tr} (gh) + {\tr}(gh^{-1})
\end{equation}
which can also be expressed graphically: for any $x \in {\ga}_{1} \cap {\ga}_{2}$: $$
{\tr}P{\exp}\oint_{{\ga}_{1}} {\CalA} \times {\tr}P{\exp}\oint_{{\ga}_{2}} {\CalA}  =
$$
\includegraphics[width=15pc]{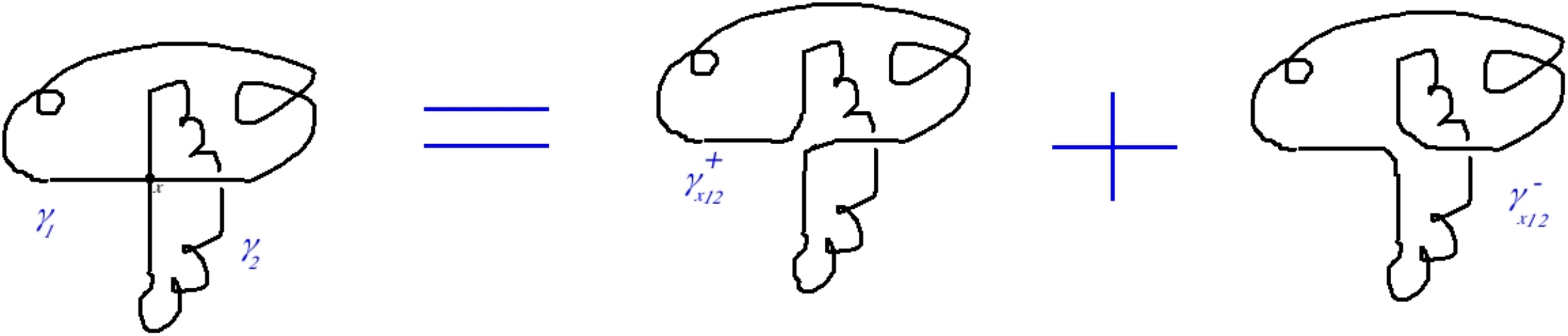}
$$ {\tr}P{\exp}\oint_{{\ga}_{x12}^{+}} {\CalA} + {\tr}P{\exp}\oint_{{\ga}_{x12}^{-}}{\CalA}
$$
We can parametrize $A, B, C$ with the help of the Darboux coordinates on ${\CalM}_{0,4; {{\n}}}$, $({\al}, {\be})$, i.e. $\{ {\al}, {\be} \} = 1$, 
\begin{equation}
\label{eq:dof}
\begin{split}
& A = e^{\al} + e^{-{\al}}\, , \\
& B (A^2 - 4)   + 2 ( m_{2}m_{3} + m_{1}m_{4} ) - A (m_{1}m_{3} + m_{2}m_{4} )\\
& \qquad =  (  e^{{\be}} + e^{-{\be}} ) \sqrt{c_{12}(A)c_{34}(A)} \, , \\
& \left( 2C + AB - m_{1}m_{3} - m_{2}m_{4} \right) \left( e^{\al} - e^{-{\al}} \right) = \\
& \qquad = \left(  e^{{\be}} - e^{-{\be}}  \right) 
\sqrt{c_{12}(A)c_{34}(A)} 
\end{split}
\end{equation}
where
$$
c_{ij}(A) = A^2 + m_{i}^2 + m_{j}^2 - A m_{i}m_{j} - 4 
$$

\subsubsection{The case of a torus with one puncture}

Now consider the case $g=1$, $n=1$. The moduli space is:
$$
{\CalM}_{1,1;{\n}} = \qquad\qquad\qquad\qquad\qquad\qquad\qquad\qquad
$$
\begin{equation}
\label{eq:modspoo}
\left\{ \left( {\gm}, {\hm} \right) \in G^{\times 2} \ \biggr\vert  
\begin{matrix}& {\tr}({\gm}{\hm}{\gm}^{-1}{\hm}^{-1}) = {m} \\
\end{matrix}
\right\} \, \biggr/ \, G 
\end{equation}
and it can be coordinatized by 
\begin{equation}
\label{eq:toh}
A = {\tr}({\gm}), \, B = {\tr} ({\hm}), \, C = {\tr} ( {\gm}{\hm} ), 
\end{equation}
which obey the relation 
\begin{equation}
\label{eq:woo}
\begin{split}
W_{1,1}(A,B,C) = 0\\
W_{1,1} = A^2 + B^2 + C^2 - ABC - m-2\\
\end{split}
\end{equation}
and the Poisson bracket of, e.g. $A$ and $B$ is easily computed to be
\begin{equation}
\left\{ A, B \right\} = C - \frac{1}{2} AB
\label{eq:abp}
\end{equation}
Now the local coordinates ${\al}, {\be}$, on ${\CalM}_{1,1;{\n}}$, s.t.
\begin{equation}
\label{eq:genone}
\begin{split}
& A = e^{\al} + e^{-{\al}}\, , \\
& B = \left( e^{{\be}/2} + e^{-{\be}/2} \right) 
\sqrt{\frac{A^2 - m -2}{A^2-4}} \\
& C = \left( e^{{\al} + {\be}/2} + e^{-{\al}-{\be}/2} \right) 
\sqrt{\frac{A^2 - m -2}{A^2-4}} 
\end{split}
\end{equation}
are the Darboux coordinates for (\ref{eq:abp}). 
Incidentally, the coordinates ${\al}, {\be}$ in (\ref{eq:genone}) are the coordinate and momentum in the two-body relativistic Calogero-Moser system with the coupling constant ${\nu}$, the so-called Rujsenaars-Schneider system, whose Hamiltonian is $B$, see \cite{Gorsky:1993dq}. 

\subsubsection{The braid and modular group actions}

The set of generators $A, B, C, \ldots $ of the ring of polynomial functions on ${\CalM}^{loc}$ depends on the choice of the generators of the fundamental group of the (punctured) surface $\Sigma$, and so do the coordinates ${\al}, {\be}$ defined by (\ref{eq:dof}), (\ref{eq:genone}). 
 For example, in the genus one case, the monodromies
 $({\gm}, {\hm})$ are defined with respect to some choice of the $A$ and $B$ cycles. An equally good choice is, e.g.
 $({\gm}, {\hm}{\gm}^{n})$, for any $n \in {\bZ}$. The corresponding $({\al}, {\be})$ coordinates transform to
 $({\al}, {\be} \pm 2 n {\al})$. There are analogous formulae in the genus zero case. We shall discuss them below. 
 
\section{The Darboux coordinates}

In this section we describe the coordinate system on the moduli space ${\CalM}_{g,n; {{\n}}}$ of flat $SL_2 ({\bC})$ connections on the punctured Riemann surface with the fixed conjugacy classes of the monodromies around the punctures. 

\subsection{The coordinate charts from the pant decomposition}

We cover ${\CalM}_{g,n; {{\n}}}$ by the coordinate charts ${\CalU}_{\Gamma}$ labelled by the points 
${\Gamma} \in {\overline{\CalM}_{g,n}}$ in the Deligne-Mumford moduli stack of stable curves of genus $g$ with $n$ punctures, corresponding to the maximally degenerate Riemann surfaces. These points are in one-to-one correspondence with the tri-valent graphs $\Gamma$ with $b_{1}({\Gamma}) = g$ with $n$ tails. Such a graph has
$3g-3+n$ internal edges (the edge is called internal if neither of its endpoints is a tail), and $2g-2+n$ internal (i.e. trivalent) vertices.   This data is equivalent to a choice of the pant decomposition of $\Sigma$. 

To each internal edge $e$ we assign a pair $({\al}_e, {\be}_e)$ of complex numbers, with some discrete identifications. The edge $e$ corresponds to a simple loop
${\gamma}_{e}$ on $\Sigma$ which gets contracted at the degeneration of the complex structure, corresponding to $\Gamma$. The holonomy around this loop determines ${\al}_{e}$ (up to the obvious indeterminancy):
\begin{equation}
{\tr} P{\exp} \oint_{{\gamma}_{e}} {\CalA}  = e^{{\al}_{e}} + e^{-{\al}_{e}}
\label{eq:hole}
\end{equation}
The rule of computing the dual coordinate ${\be}_{e}$ is the following. 
There are two situations: the two endpoints of the edge $e$ are disctinct, in which case we call $e$  {\em the genus $0$ edge}, 
\begin{figure}
\includegraphics[height=8pc]{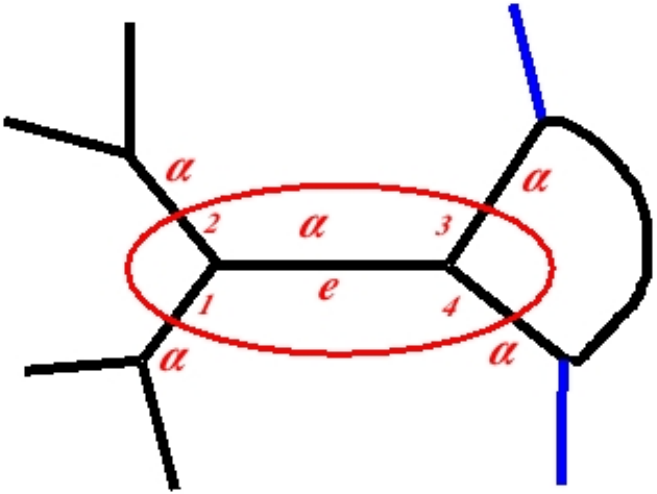}\\
\qquad \includegraphics[height=8pc]{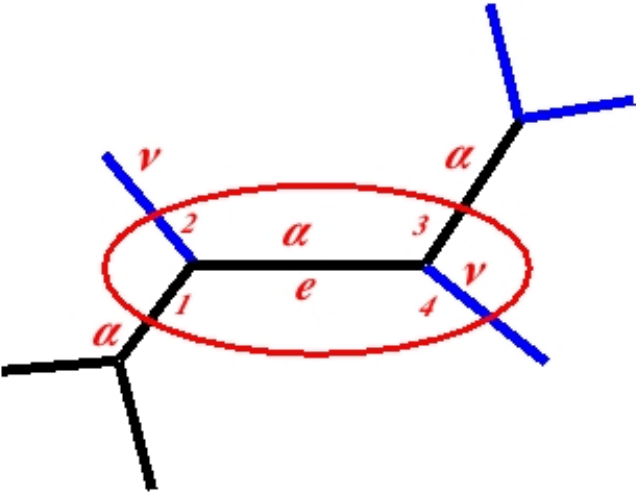}\\
\qquad\qquad \includegraphics[height=8pc]{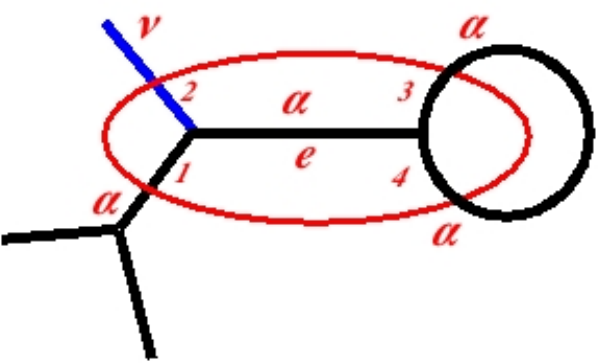}
\caption{The genus $0$ edges}
\label{fig:g0c}
\end{figure}
or the two endpoints coincide, in which case we call $e$ {\em the genus $1$ edge.} 
\begin{figure}
\includegraphics[height=5pc]{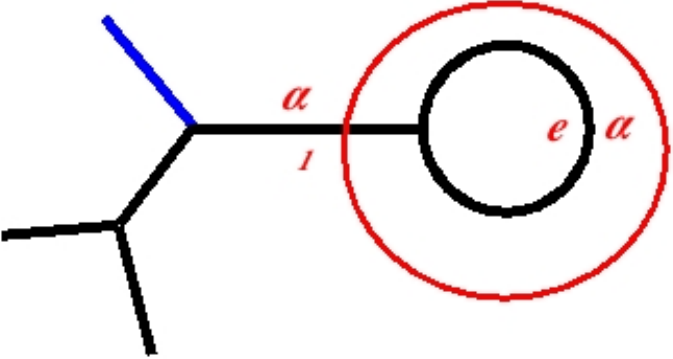}\\
\qquad \includegraphics[height=5pc]{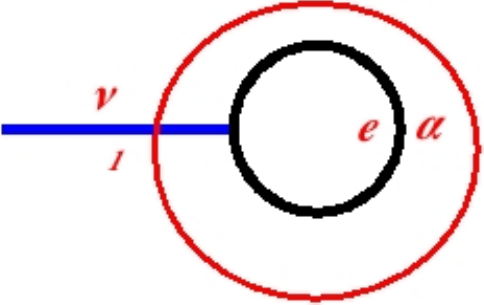}
\caption{The genus $1$ edges}
\label{fig:g1c}
\end{figure}
Each of the two endpoints of the genus $0$ edge $e$ has two other edges emanating, some of them could be tails, some of them could be identical, see the Figs. \ref{fig:g0c}.  By cutting open the simple loops on $\Sigma$ corresponding to the edges emanating
out of the two end-points of $e$, we get a sphere with four holes, which is cut out of $\Sigma$. 
Let us enumerate these holes, so that the two holes on the one end are $1$ and $2$, and the other two are $3$ and $4$. The flat connection on $\Sigma$ restricts to the flat connection on the four-holed sphere, and defines the invariant functions, $A$, $B$, $C$, as in
(\ref{eq:genfr}). Then, ${\al}$ and ${\be}$ defined by
(\ref{eq:dof}) are the ${\al}_{e}$ and ${\be}_{e}$, corresponding to the edge $e$.
We could have mislabeled the two edges emanating  from one end of $e$. This would have replaced ${\be}_{e}$
by ${\be}_{e} \pm {\al}_{e}$. 

For the genus $1$ edge the situation is similar, except that there is only one simple loop to cut, which extracts out of $\Sigma$ a one-holed torus. The restriction of flat connection onto this torus defines the invariant functions $A, B$ and $C$ as in (\ref{eq:toh}). 
Then, using (\ref{eq:genone}) we define the $({\al}_{e}, {\be}_{e})$ pair for the local genus $1$ edge.

\subsubsection{A little bit of geometry}

In this section we shall review some standard facts about the geometry of symplectic surfaces, the complexification of the Euclidean and Lobachevsky geometries, and the spaces of polygons.

We shall construct a complex version of the spherical and hyperbolic geometry of polygons in $M^3 = {\bf S}^{3}$ or 
$M^3 = {\bH}^{3}$. It turns out that the Darboux coordinates ${\al}$ and ${\be}$ of the previous considerations can be viewed as describing the geometry of quadrangles in the $M^3_{\bC} \approx G$. The justification for the somewhat involved analysis of the simple geometry is that the generalization of the Darboux coordinates for $n > 4$ turns out to describe the $n$-gons in $G$.

Let us view the group $G$ as the affine hypersurface in the four dimensional complex vector space $V = {\bC}^4 = {\rm Mat}_{2}({\bC})$ of the $2 \times 2$ complex
matrices. We endow  $V$ with the complex metric:
\begin{equation}
\label{eq:cmpl}
\langle X , Y \rangle = \frac{1}{2} \left( {\tr}(XY) - {\tr}X {\tr}Y \right)
\end{equation}
which is clearly invariant under the action of the group $G \times G$:
\begin{equation}
\label{eq:ggx}
X \mapsto (g_{L}, g_{R}) \cdot X = g_{L} X g_{R}^{-1}
\end{equation}
 The group $G$ is realized as a hypersurface:
 \begin{equation}
 \label{eq:grg}
 \langle g, g \rangle = 1 \, \Leftrightarrow \, g \in G
 \end{equation}
The metric (\ref{eq:cmpl}) induces a 
complex metric on $G$:
\begin{equation}
\label{eq:cmplmg}
ds^2_{G} = \frac{1}{2} {\tr}dg^2- \frac{1}{2} ({\tr}dg)^2   = - \frac{1}{2} {\tr} \left( g^{-1}dg \right)^2
\end{equation}
 where we used the identity in $G \subset {\rm Mat}_{2\times 2}({\bC})$:
 \begin{equation}
 \label{eq:sltid}
 g+g^{-1} = {\tr}g \cdot 1
 \end{equation}
The volume form $dg_{11}dg_{12}dg_{21}dg_{22}$ on $V$
induces the volume form vol$_{G} = {1\over 8{\pi}^{2}}{\tr} \left( g^{-1}dg \right)^{3}$ on $G$ which is normalized so as to have a period $1$ on the compact
three-cycle ${\bf S}^{3} \subset G$. 
 
Given four elements ${\gm}_{1}, {\gm}_{2}, {\gm}_{3}, {\gm}_{4} \in G$, which obey ${\gm}_{4}{\gm}_{3}{\gm}_{2}{\gm}_{1} = 1$, we construct the complexified
analogue of the hyperbolic tetrahedron $\Delta$ (see the Figure) . The vertices of $\Delta$ are the points: $(v_{1} = 1, v_{2} = {\gm}_{1}, v_{3}={\gm}_{2}{\gm}_{1}, v_{4} = {\gm}_{3}{\gm}_{2}{\gm}_{1})$. Let us introduce the 
Gram matrix of $\Delta$:
\begin{equation}
\label{eq:grm}
{\CalC} = \Vert c_{ij}  \Vert_{i,j=1\ldots 4}, \qquad c_{ij} = c_{ji} = \langle v_{i}, v_{j} \rangle
\end{equation}
We shall assume this matrix non-degenerate, ${\rm Det}({\CalC}) \neq 0$. This implies that the vectors $v_{1}, \ldots , v_{4}$ are linearly independent.  In particular, since $\langle v_i , v_i \rangle
 = 1$ for any $i$, we have that $c_{ij} \neq \pm 1$ for any 
 $i\neq j$.
Let us introduce the (hyperbolic) angle ${\al}_{ij}$, via
\begin{equation}
\label{eq:hypa}
{\rm cosh}({\al}_{ij}) = c_{ij}
\end{equation}

\includegraphics[width=15pc]{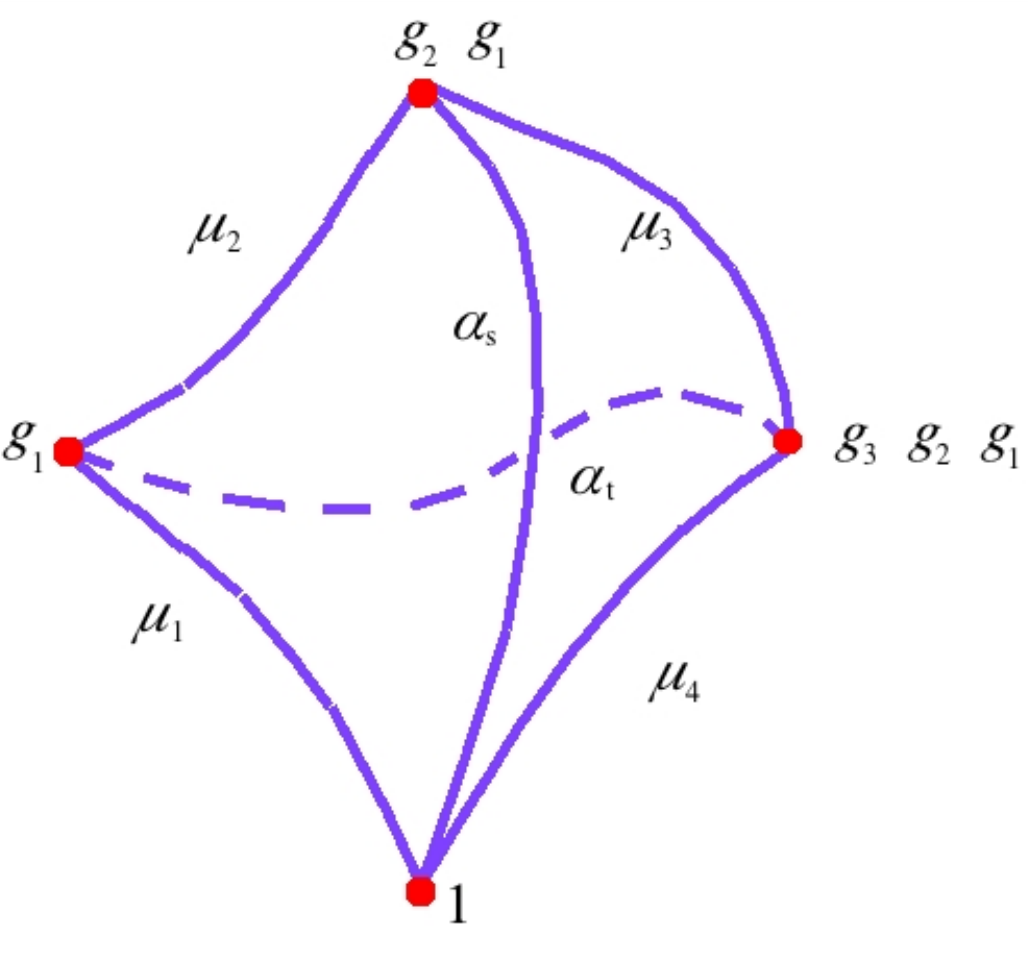}

The choice of ${\al}_{ij}$ given $c_{ij}$ is of course not unique, even up to a shift by $2{\pi}i{\bZ}$. To fix the choice of ${\al}_{ij}$ as a unique element in ${\bC}/2{\pi}i{\bZ}$, one can pick 
$s_{ij} = {\rm sinh}({\al}_{ij})$ in addition to $c_{ij}$.     
The edge $e_{ij}$ connecting the vertices $v_{i}$ and $v_{j}$ is the intersection of the two dimensional complex
plane $E_{ij} = {\rm Span}( v_{i}, v_{j} ) \subset V$ and $G$:
$$
e_{ij} = E_{ij} \cap G =
$$
\begin{equation}
\label{eq:edgij}
 \{ t_{1} v_{i}
+ t_{2} v_{j} \, |\, t_{1}^2 + t_{2}^2 + 2t_{1}t_{2} c_{ij} = 1, \, t_{1},t_{2} \in {\bC} \, \}
\end{equation} 
We can identify $e_{ij}$ with a copy of ${\bC}^{\times}$, using the explicit parametrization: for $z \in {\bC}^{\times}$, 
\begin{equation}
\label{eq:parame}
t_{1} = \frac{ e^{ 2{\al}_{ij}} - z^2}{z ( e^{2{\al}_{ij}} - 1 ) }  
\end{equation}
$$ 
t_{2} = \frac{ e^{{\al}_{ij}} ( z^2 - 1 )}{z ( e^{2{\al}_{ij}} - 1 )}
$$
The metric (\ref{eq:cmplmg}) restricts to $e_{ij}$, as
\begin{equation}
\label{eq:dse}
ds^2_{e_{ij}}  = - \left( {dz\over z} \right)^2 \ . 
\end{equation} 
Thus, on $e_{ij}$, the complex metric 
(\ref{eq:dse}) is a square of a $(1,0)$-differential $ds = i dz/z$. The "length" of the edge $l_{ij}$ is the period of $ds$ along the one-chain connecting the vertices $v_{i}$ and $v_{j}$, which in the $z$-coordinate is a path ${\CalC}_{ij}$ connecting the points $1$ and $e^{{\al}_{ij}}$. 
\begin{equation}
\label{eq:lngth}
\int_{{\CalC}_{ij}} ds = {\al}_{ij} \, {\rm mod} \, 2{\pi}i{\bZ}
\end{equation}
The face $f_{ijk}$ of the the tetrahedron, with the vertices $v_{i}, v_{j}, v_{k}$, is the intersection of the three complex dimensional vector space $F_{ijk} = {\rm Span}(v_{i}, v_{j}, v_{k}) \subset V$ and $G$:
$$
f_{ijk} = F_{ijk} \cap G = \qquad
$$
\begin{equation}
\label{eq:faceij}
\{ \, \ t_{1} v_{i}
+ t_{2} v_{j} + t_{3} v_{k} \, |\, t_{1},t_{2},t_{3} \in {\bC}  , \ \end{equation}
$$t_{1}^2 + t_{2}^2 + t_{3}^{2} + 2t_{1}t_{2} c_{ij} + 2t_{1}t_{3} c_{ik} + 2t_{2}t_{3} c_{jk} = 1\, \}
$$
The manifold $f_{ijk}$ is diffeomorphic to $T^{*}{\bf S}^{2}$. 
Given two faces, $f_{ijk}$ and $f_{jkl}$, with the common
edge $l_{jk}$, we define the angle ${\be}_{il}$ between them, as the angle between the planes $F_{ijk}$ and $F_{jkl}$. The latter is defined as the angle between the normal lines, $n_{ijk}$ and $n_{jkl}$, once a coorientation of the faces $f_{ijk}$ and $f_{jkl}$ is picked. Let us explain what it is. 

The normal line $n$ to a three-dimensional vector subspace $F \subset V$ is
defined as the one-dimensional complex line in $V$, $n \subset V$, orthogonal, in the sense of the metric (\ref{eq:cmpl}),
to $F$: for any $f \in F$, $\langle f, n \rangle = 0$. 
There are two vectors of unit norm in $n$, which differ by a sign. A choice of one of these two unit norm vectors is the choice of the coorientation of $F$ in $V$. 
We shall denote the unit norm vector in $n$ by the same 
letter $n$. 

Once the coorientation is fixed, the angle is defined via:
\begin{equation}
\label{eq:angfac}
{\rm cosh}({\be}_{il}) = \langle n_{ijk}, n_{jkl} \rangle
\end{equation}
A change of coorientation of one of the two planes changes the angle ${\be}$ to ${\pi}i - {\be}$. This ambiguity is in addition to the ambiguity ${\be} \mapsto - {\be}$, and ${\be} \mapsto {\be} + 2{\pi} ik$, $k \in {\bZ}$. 

Let us now make the following useful observation (cf. \cite{volumehyper}):
Let $^{L}{\CalC}$ be the matrix of the hyperbolic cosines of the dihedral angles ${\be}_{ij}$:
\begin{equation}
\label{eq:lcc} ^{L}{\CalC} = \Vert {\rm cosh}( {\be}_{ij} ) \Vert_{i,j=1}^{4}
\end{equation} 
Then:
\begin{equation}
\label{eq:cclcc}
^{L}{\CalC} = {1\over \sqrt{{\CalC}^{\vee}_{d}}}\
{\CalC}^{\vee} \, {1\over \sqrt{{\CalC}^{\vee}_{d}}}
\end{equation}
where ${\CalC}^{\vee} = {\rm Det}({\CalC}) {\CalC}^{-1}$ is
the matrix of the minors of $\CalC$, and ${\CalC}^{\vee}_{d}$ is its diagonal part, e.g.
\begin{equation}
\label{eq:coshb}
{\rm cosh}({\be}_{ij}) = {{\, {\CalC}^{\vee}_{ij} }  \over \sqrt{{\CalC}^{\vee}_{ii}{\CalC}^{\vee}_{jj}}}
\end{equation}
The coorientation ambiguity in the definition of the $\be$
angles is the reason for the square root ambiguity in (\ref{eq:coshb}). 
To demonstrate (\ref{eq:coshb}) let us note that the vectors
\begin{equation}
\label{eq:vvee}
v_{i}^{\vee} = \sum_{j=1}^{4} {\CalC}_{ij}^{\vee}  v_{j}
\end{equation}
obey
\begin{equation}
\label{eq:cvee}
\langle v_{i}^{\vee} , v_{k} \rangle =
{\rm Det}({\CalC}) {\de}_{ik}
\end{equation}
and
\begin{equation}
\label{eq:cveesc}
\langle v_{i}^{\vee}, v_{j}^{\vee} \rangle = 
{\rm Det}({\CalC}) {\CalC}^{\vee}_{ij}
\end{equation}
Therefore, we can choose the normal vectors to be
\begin{equation}
\label{eq:nrmvc}
n_{ijk} = {\ve}_{ijkl} {v_{l}^{\vee} \over \sqrt{{\rm Det}({\CalC}) {\CalC}^{\vee}_{ll}}}
\end{equation}
from which (\ref{eq:coshb}) follows immediately. 

 We can interpret the matrix $^{L}{\CalC}$ as the Gram matrix of the dual tetrahedron 
 $$
 ^{L}{\Delta}\quad  \subset\quad  
  ^{L}G = PGL(2, {\bC})$$ whose vertices are the normals to the faces of ${\Delta}$ up to a choice of orientation.  
 Now let us return to the study of the moduli space 
 ${\CalM}_{0,4;\vec \n}$. We assign to every point in 
 a finite cover of ${\CalM}_{0,4;{\n}}$ a tetrahedron (perhaps, degenerate) in $G$, up to the action of the group $G \times G$ of the isometries of the metric (\ref{eq:cmpl}). The vertices of the tetrahedron can be chosen to be:
 \begin{equation}
 \label{eq:verts}
 (v_1, v_2, v_3, v_4) = (1, {\gm}_1, {\gm}_2 {\gm}_1, {\gm}_3 {\gm}_2 {\gm}_1 )\end{equation}
The corresponding Gram matrix is readily computed:
\begin{equation}
\label{eq:gramgr}
{\CalC} = {1\over 2} \left( \begin{matrix}
2 & {m}_1 & A & m_4 \\
m_1 & 2 & m_2 & B \\
A & m_2 & 2 & m_3 \\
m_4 & B & m_3 & 2 \\ \end{matrix} \right) 
\end{equation} 
which gives 
$$
-4{\CalC}^{\vee}_{22} = c_{34}(A), \ -4{\CalC}^{\vee}_{44} = c_{12}(A), \ 
$$
\begin{equation}
\label{eq:dlgram}
8 {\CalC}^{\vee}_{24} =  B (A^2-4) - A (m_{1}m_{3}+m_{2}m_{4})  \end{equation}
$$
\qquad\qquad\qquad\qquad +2(m_2 m_3 + m_1 m_4 )
$$

\subsubsection{The Darboux coordinates for ${\CalM}_{0,4;{{\n}}}$}
\begin{figure}
 \includegraphics[width=15pc]{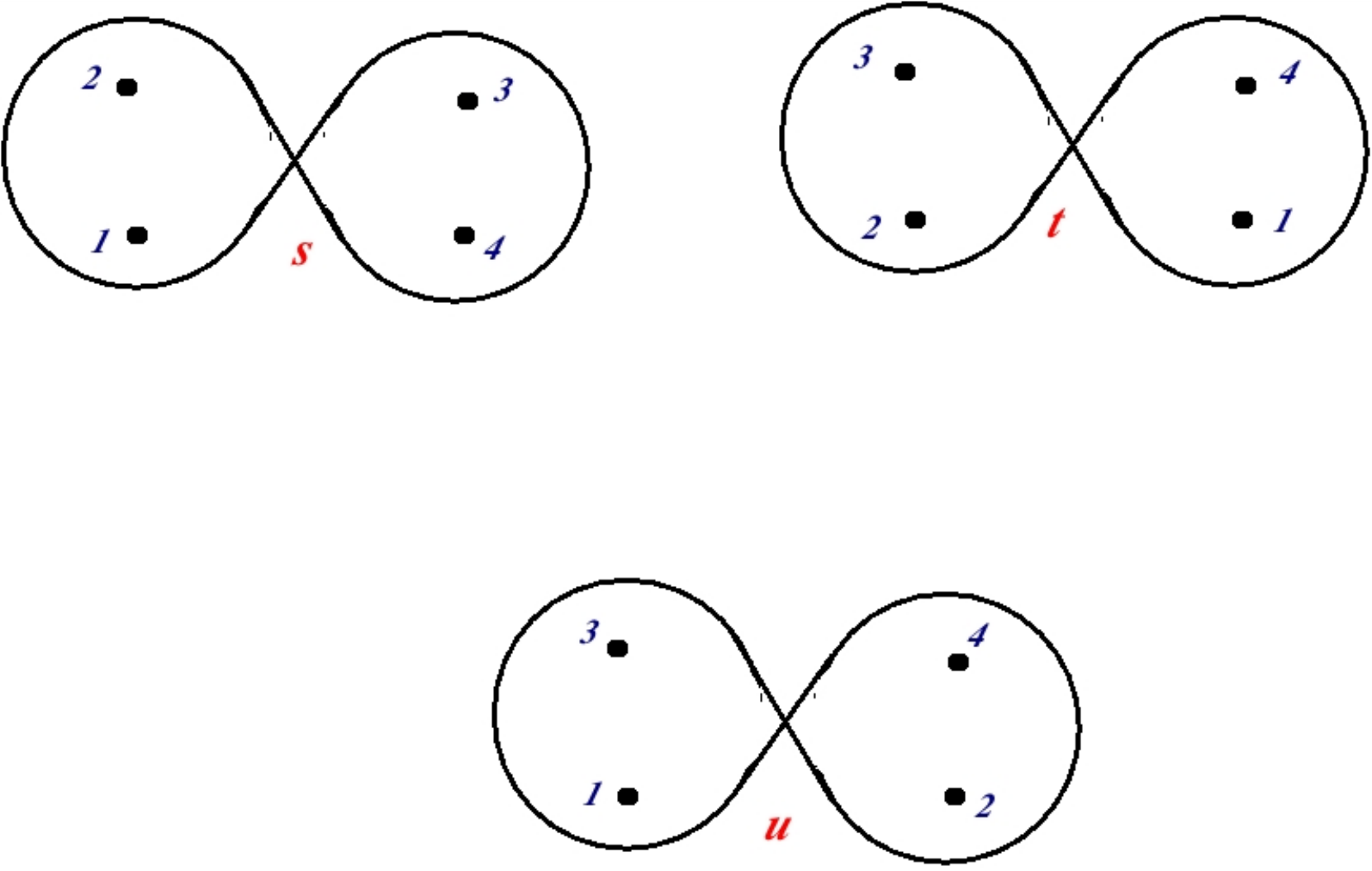}
 \caption{The degeneration points in ${\overline {\CalM}_{0,4}}$}
 \label{fig:stu}
 \end{figure}

 We are now in position to state the definition of our coordinates in the case of the $4$-punctured sphere. 
 There are three points of the maximal degeneration of the complex structure which correspond to the $s$, $t$, and $u$ channel tree scattering graphs. These degenerations collide the points $1-2, 3-4$,
 the points $1-4, 2-3$ and the points $1-3, 2-4$, respectively, see the Fig. \ref{fig:stu}.
We cover the moduli space of flat connections by three coordinate charts. 
Each chart ${\CalU}_i$ has the coordinates $({\al}_i, {\be}_i)$, where $i = s, t, u$. The relation of $({\al}_{s},{\be}_{s})$, and $({\al}_{t}, {\be}_{t})$ coordinates to the flat connection on the $4$-punctured sphere with the basic holonomies ${\gm}_{1,2,3}$ is depicted in the 
\begin{figure}
\includegraphics[width=15pc]{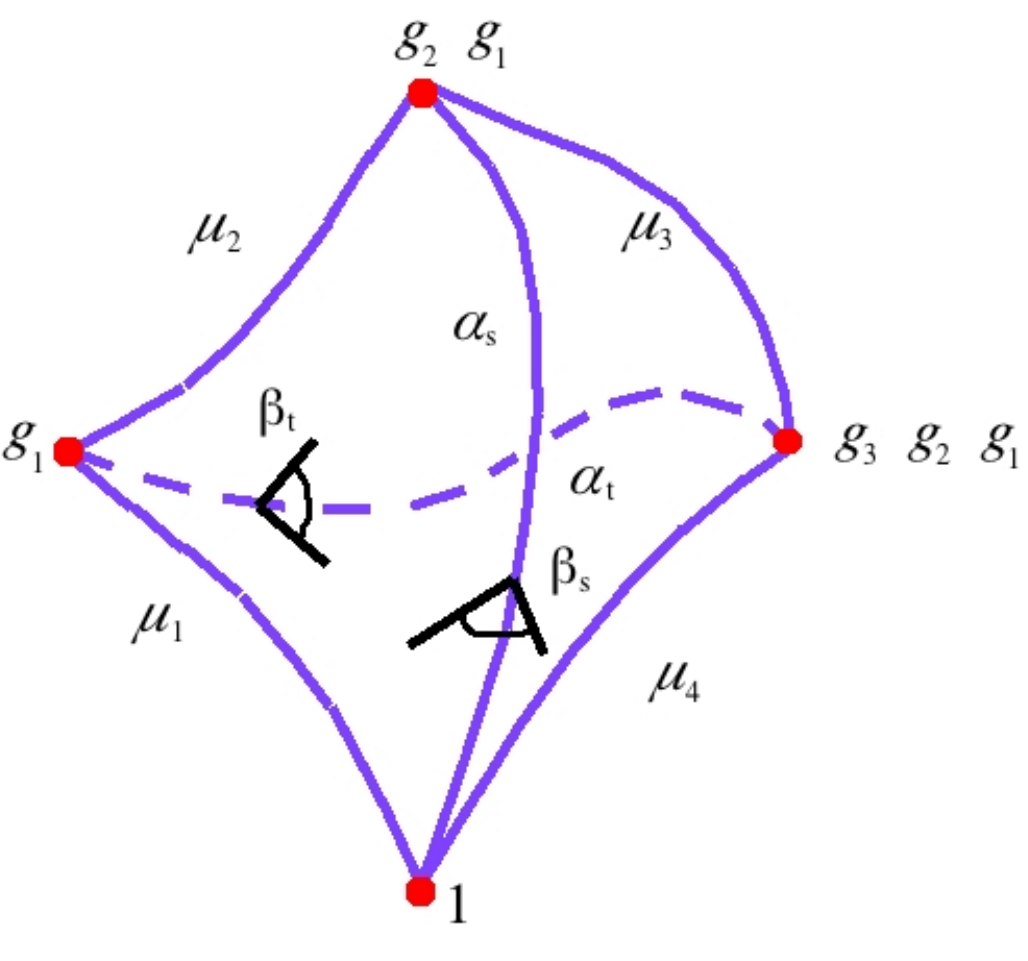}
\caption{The $({\al}_{s}, {\be}_{s})$ and $({\al}_{t}, {\be}_{t})$ coordinates}
\label{fig:dihang}
\end{figure}
Fig. \ref{fig:dihang}. The coordinates $({\al}_{u}, {\be}_{u})$ are obtained by applying the modular group action:
\begin{equation}
\label{eq:modg}
\left( {\gm}_{1}, {\gm}_{2}, {\gm}_{3} \right) \mapsto
\left( {\gm}_{2}, {\gm}_{2}{\gm}_{1}{\gm}_{2}^{-1}, {\gm}_{3} \right)
\end{equation}
which maps the tetrahedron with the vertices
$(1, {\gm}_{1}, {\gm}_{2}{\gm}_{1}, {\gm}_{3}{\gm}_{2}{\gm}_{1})$ to the tetrahedron with the vertices
$(1, {\gm}_{2}, {\gm}_{2}{\gm}_{1}, {\gm}_{3}{\gm}_{2}{\gm}_{1})$.

\subsubsection{The Darboux coordinates for ${\CalM}_{0,n; {{\n}}}$}

are defined analogously, and can be given a geometric interpretation in terms of the geometry of a polygon
with the vertices $(1, {\gm}_{1}, {\gm}_{2}{\gm}_{1}, \ldots, {\gm}_{n-1}{\gm}_{n-2}\ldots {\gm}_{2}{\gm}_{1} )$ which is cut by $n-3$ diagonals onto $n-2$ (complexified hyperbolic or spherical) triangles. 
The coordinate ${\al}_{i}$, $i = 1, \ldots, n-3$ is the
hyperbolic length of the diagonal $d_{i}$, while the coordinate
${\be}_{i}$ is the dihedral angle between the two adjacent triangles, which share the common edge $d_{i}$, see 
\begin{figure}
\includegraphics[width=15pc]{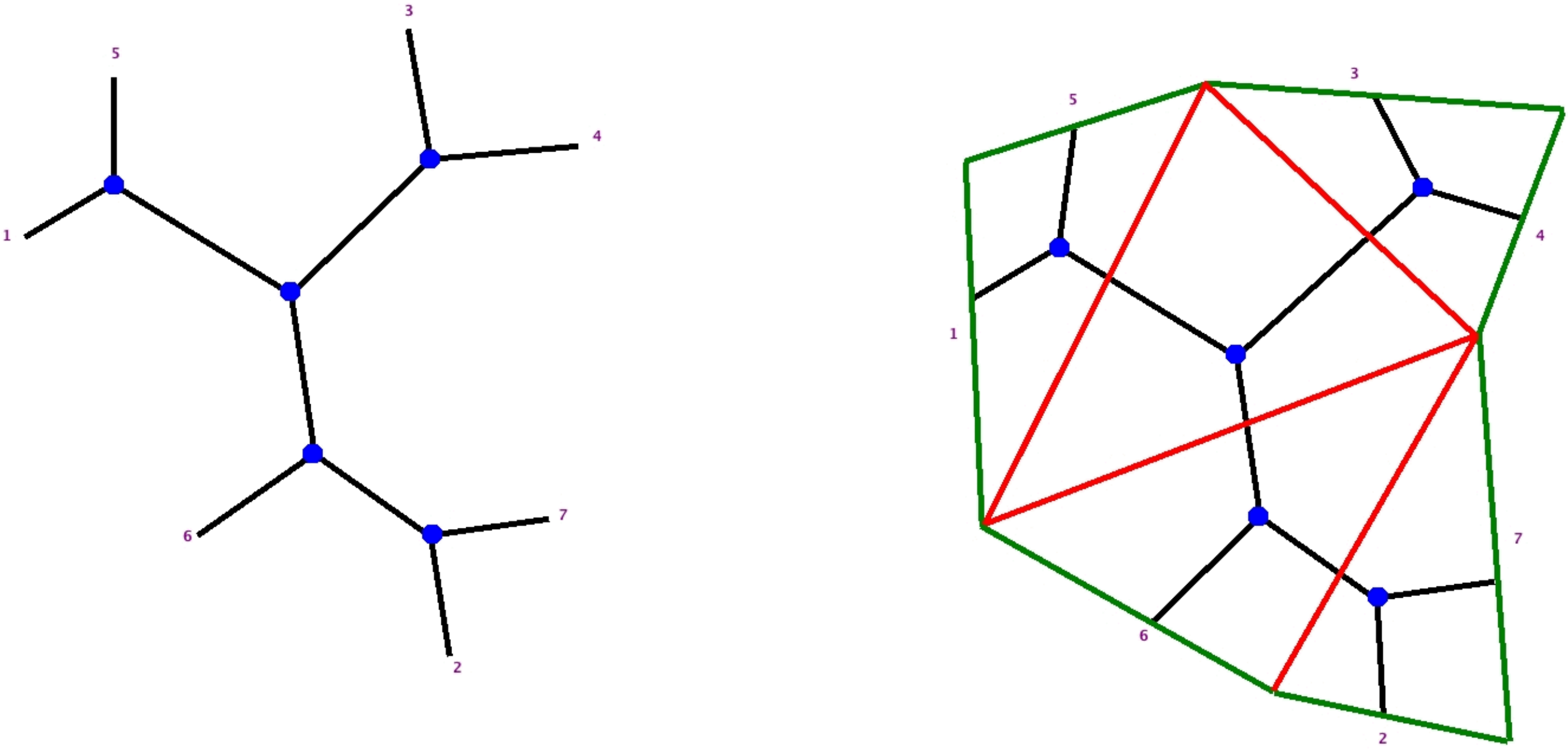}
\caption{From the tree graph to the polygon triangulated by the diagonals} 
\label{fig:trgrdg}
\end{figure}
Fig. \ref{fig:trgrdg} for the case of $n=7$. 

\subsubsection{The coordinate transformations}

It is interesting to look at the coordinate transformations which glue our coordinate systems on the overlaps ${\CalU}_{\Gamma} \cap {\CalU}_{{\Gamma}'}$. 
It suffices to discuss the overlaps where the two graphs
${\Gamma}$ and ${\Gamma}'$ differ in exactly one genus $0$ edge, or in exactly one genus $1$ edge. In the first case the basic transformation is 
$({\al}_{s}, {\be}_{s}) \mapsto ({\al}_{t}, {\be}_{t})$
in the case of ${\CalM}_{0,4; {{\n}}}$. Recall that the canonical transformations, i.e. the transformations
preserving the symplectic form, are generated by a function, the so-called generating function:
${\CalS}_{0,4}({\al}_{s}, {\al}_{t}; {\vec \n})$, 
such that:
\begin{equation}
\label{eq:genfuncs}
{\be}_{s} = \frac{{\p}{\CalS}_{0,4}}{{\p}{\al}_{s}} \, , \quad
 {\be}_{t} = \frac{{\p}{\CalS}_{0,4}}{{\p}{\al}_{t}} 
\end{equation}
Using the formulae of \cite{schlafli} one can easily derive
that
${\CalS}_{0,4}({\al}_{s}, {\al}_{t}, {\n}_{1}, {\n}_{2}, {\n}_{3}, {\n}_{4})$ is the hyperbolic volume of the  
tetrahedron $^{L}{\Delta}$, which is dual to the tetrahedron ${\Delta}$ (Fig. \ref{fig:dihang}) whose edges have the lengths 
$$
{\al}_{s}, {\al}_{t}, {\m}_{1}, {\m}_{2}, {\m}_{3}, {\m}_{4}\ .
$$
where ${\m}_{k} = 2\pi i {\n}_{k}$. 
The explicit formula can be derived using 
\cite{volumehyper} (see also \cite{schlafli}, \cite{lobach}, \cite{volcho}, \cite{volmilnor}): Let
\begin{multline*}
 {\CalV}_{1} = {\al}_{s} + {\m}_{1} + {\m}_{2} \\
 {\CalV}_{2} = {\al}_{s} + {\m}_{3} + {\m}_{4} \\
 {\CalV}_{3} = {\al}_{t} + {\m}_{1} + {\m}_{4} \\
 {\CalV}_{4} = {\al}_{t} + {\m}_{2} + {\m}_{3} \\
 {\CalH}_{1} = {\m}_{1} + {\m}_{2} + {\m}_{3} + {\m}_{4} \\
 {\CalH}_{2} = {\al}_{s}+ {\al}_{t} + {\m}_{1}+{\m}_{3} \\
 {\CalH}_{3} = {\al}_{s}+{\al}_{t} + {\m}_{2} + {\m}_{4}\\
 {\CalH}_{4} = 0 \\
\end{multline*} 
Then:
\begin{equation}
\label{eq:csvol}
\begin{split}
{\CalS}_{0,4}({\al}_{s}, {\al}_{t} , {\vec \n} ) = \qquad\qquad\qquad\qquad \\
\frac{1}{2} \sum_{a=1}^{4} \left[ Li_{2} \left( w_{+} e^{{\CalV}_{a}} \right) -
Li_{2} \left( w_{-} e^{{\CalV}_{a}} \right) \right. \\
 \qquad\qquad\left.   - 
Li_{2} \left( w_{+} e^{{\CalH}_{a}} \right) 
+
Li_{2} \left( w_{-} e^{{\CalH}_{a}} \right) \right] \\
\end{split}
\end{equation}
where
$w_{\pm}$ are the two different roots of the (quadratic in $w$)
equation:
\begin{equation}
\label{eq:eqzz}
\prod_{a=1}^{4} (1 - w e^{{\CalV}_{a}}) = \prod_{a=1}^{4} ( 1 - w e^{{\CalH}_{a}})
\end{equation}
For the genus $1$ edge the formalism is similar and we shall not present it here. 

\subsubsection{The Hamiltonian flows and bending}

Fix $\Gamma$ and consider the Hamiltonian flows generated by the Poisson-commuting functions 
$A_{e} = e^{{\al}_{e}} + e^{-{\al}_{e}}$, for all edges $e$. These flows generate a complexified integrable system. 
Its real slices include the Fenchel-Nielsen twistings on the Teichm\"uller space and Goldman flows \cite{Tyurin} on the moduli space of $SU(2)$ flat connections.

In our coordinates the flow generated by ${\al}_{e}$ acts very simply: ${\be}_{e}$ is shifted while ${\al}_{e'}$'s and 
${\be}_{e'}$ for $e' \neq e$ are unchanged. It is amusing to see the action of this flow on the monodromy data of a flat connection ${\CalA}$. We shall do it for the $g=0, n=4$ case. 

Thus, let us take ${\al}$ as a Hamiltonian, and let us try to find out what is the Hamiltonian flow ${\CalU}_{t}$ generated by it. Of course, 
in the $({\al}, {\be})$ coordinates this is trivial: ${\CalU}_{t}: ({\al}, {\be}) \mapsto ({\al} , {\be}+ t)$. 
Let us calculate the effect of this transformation on the flat connection ${\CalA}$, which we parametrize by the gauge equivalence class of the triple $({\gm}_{1}, {\gm}_{2}, {\gm}_{3}, {\gm}_4)$, with ${\gm}_4{\gm}_3 {\gm}_2{\gm}_1 = 1$. 
We claim:
\begin{equation}
\label{eq:ggggt} 
{\CalU}_{t}: ({\gm}_{1}, {\gm}_{2}, {\gm}_{3}, {\gm}_{4} ) \mapsto  ({\gm}_{1}, {\gm}_{2}, e^{- tJ} {\gm}_{3} e^{ tJ}, e^{- tJ}{\gm}_{4}e^{ tJ} )
\end{equation}
where we have introduced a normalized Lie algebra element
$J$, a traceless $2 \times 2$ matrix with the eigenvalues $\pm \frac{1}{2}$, 
${\tr}J = 0, \qquad {\tr}J^2 = {1\over 2}$, 
such that:
\begin{equation}
\label{eq:jgg}
{\gm}_{2}{\gm}_{1} = e^{2{\al}J}\ .
\end{equation}
Indeed, let 
\begin{equation}
\label{eq:btdf}
B_{t} = {\tr}({\gm}_{2}e^{- tJ}{\gm}_{3}e^{ tJ})
\end{equation}
Using the identity for the $SL_2$ matrices:
$$
e^{2t{\al} J} = {{\rm sinh}\, (1-t) {\al} \over
{\rm sinh} \, {\al}  }  + {{\rm sinh} \, t {\al} \over {\rm sinh}\, {\al}} \, e^{2{\al}J} \, .
$$
we can easily compute:
\begin{equation*}
\begin{split}
B_{t}(A^2-4) + 2\,  (m_{2}m_{3}+m_{1}m_{4}) - \\
A  \, (m_{1}m_{3}+m_{2}m_{4})  = \\  2\, {\rm cosh}({\be}+ t) \sqrt{c_{12}(A) c_{34}(A)} \\
\end{split}
\end{equation*}
which establishes (\ref{eq:ggggt}). 
 
In the general $g=0$ case, the Hamiltonian flow generated by ${\al}_{e}$ corresponding to the edge $e$, and 
the corresponding diagonal $d_{e}$ (as in the Fig. \ref{fig:trgrdg}), 
has a very simple geometric interpretation (modulo the complexification): one simply bends the hyperbolic $n$-gon along the diagonal $d_{e}$. The angle of bending corresponding to ${\CalU}_{t}$ is equal to $t$. This is a complexified version of the constructions in \cite{Goldman:1986if}, \cite{Kapovich:1999math}, \cite{Klyachko}, \cite{Foth:2007}.

\section{The brane of opers}

In this section we study briefly the so-called brane of opers in the sigma model with the target space ${\CalM}^{loc}_{\Sigma}$. This is a $(A, B, A)$ type D-brane ${\CalB}_{{\CalO}_{\tau}}$, which corresponds to the $J$-complex ${\Omega}_{J}$-Lagrangian submanifold ${\CalO}_{\tau} \subset {\CalM}^{loc}_{\Sigma}$, described in \cite{BeDr}. 
In fact, locally ${\CalM}^{loc}$ is foliated by the varieties of $SL_2$-opers for different complex structures on $\Sigma$, as expressed by the local identification of the moduli space of flat $SL_2$-connections with the moduli space of projective structures \cite{Bilal:1990}, see also
\cite{Verlinde:1990h}, \cite{Gerasimov:1990fs}. 

Roughly speaking, a flat connection ${\CalA} = {\CalA}_{z}dz + {\CalA}_{\bar z} d{\bar z}$ is a $G$-oper, if the gauge equivalence class of ${\bar A} = {\CalA}_{\bar z}d{\bar z}$
defines a particular holomorphic $G$-bundle on $\Sigma$, which is determined by the complex structure of $\Sigma$. For $G = SL_{2}({\bC})$ this bundle is such that the associated rank two vector bundle is the (unique up to isomorphism) nontrivial extension of the bundle $K_{\Sigma}^{-1/2}$ by $K_{\Sigma}^{1/2}$.

Locally an $SL_2$-oper is a second order (meromorphic) differential operator 
${\CalD} = - {\p}^2 + T(z)$
which acts on the $( -\frac{1}{2})$-differentials. 

If we fix for convenience some reference complex structure on $\Sigma$, with the local coordinates $(w,{\bar w})$, and describe the generic complex structure
with the help of the Beltrami differential ${\mu} = {\mu}_{\bar w}^{w} d{\bar w} {\p}_{w}$, then 
\begin{equation}
\label{eq:awawb}
A_{\bar w} - {\mu} A_{w} = \left( \begin{matrix}
- \frac{1}{2} {\p}{\mu} & 0 \\
- \frac{1}{2} {\p}^{2} {\mu} & \frac{1}{2} {\p}{\mu} \end{matrix} \right) , \ 
A_{w} = \left( \begin{matrix}
0 & 1 \\
{\tilde T} & 0 \end{matrix} \right) 
\end{equation}
where ${\tilde T}$ obeys the following compatibility condition 
\begin{equation}
\left( {\bar\p} - {\mu}{\p} - 2{\p}{\mu} \right) {\tilde T} = - \frac{1}{2}{\p}^{3} {\mu}
\label{eq:compat}
\end{equation}
The notion of a $G$-oper  was in general given for a Riemann surface with punctures, where the opers may develop certain poles \cite{BeDr}. In this paper we study only the case of regular singularities. It means at any puncture, $T$ has at most the second order pole. 

In the case of $G = SL_2({\bC})$ the space of opers for varying complex structure of $\Sigma$ is the open subset in the moduli space of flat $G$-connections \cite{Bilal:1990}. 

\subsection{The $SL_2$-opers in genus zero}

In the case of the genus zero surface with $n$ marked points the $SL_2$-oper can be identified with the second order differential operator (the projective connection)
of the form:
\begin{equation}
\label{eq:gzoper}
\begin{split}
{\CalD} = -{\p}^{2}_{z} + T(z) \qquad\qquad\qquad\qquad \\
T(z) = \sum_{a=1}^{n}\ \frac{{\Delta}_{a}}{(z-x_{a})^2} + \frac{{\ep}_a}{z - x_a} \\
\end{split}
\end{equation}
where the {\it accessory parameters} ${\ep}_{a}$ obey
\begin{equation}
\label{eq:thmm}
\begin{split}
\sum_{a=1}^{n} {\ep}_{a} = 0 \\
\sum_{a=1}^{n} \left( x_{a}{\ep}_{a} + {\Delta}_{a} \right) = 0 \\
\sum_{a=1}^{n} \left( x_{a}^2 {\ep}_{a} + 2x_{a}{\Delta}_{a} \right) = 0 
\end{split}
\end{equation}
in order for (\ref{eq:gzoper}) be non-singular at $z = \infty$. 
Actually, an open set in the space of all $SL_2$-flat
connections on the $n$-punctured sphere can be parametrized by the space of $n$-tuples
$(x_{a}, {\ep}_{a})_{a=1}^{n}$ obeying (\ref{eq:thmm}) modulo the diagonal
$PGL_2$-action
$$
\left( x_{a}, {\ep}_{a} \right)_{a=1}^{n} \mapsto 
$$
$$
\left( {A x_a + B \over C x_a + D} ,\ {\ep}_{a} ( C x_{a} + D )^2 + 2 C ( C x_a + D ) {\Delta}_a \right)_{a=1}^{n}
$$
The oper (\ref{eq:gzoper}) defines a point in ${\CalM}_{0,n; {{\n}}}$ when
\begin{equation}
\label{eq:denu}
{\Delta}_{a} = {\n}_{a}({\n}_{a}-1)\, , \ a = 1, \ldots n 
\end{equation}
The correspondence (\ref{eq:belt},\ref{eq:hamh}) ${\p}_{{\tau}_{k}} \leftrightarrow H_k$ between the variations of the complex moduli of $\Sigma$ and the functions on the variety of opers ${\CalO}_{\tau}$ is provided, in this case,
by the one-form
\begin{equation}
{\de} = \sum_{a} {\ep}_{a} dx_{a}
\label{eq:corrsgz}
\end{equation}
Once a global coordinate $z$ on the sphere is fixed,
e.g. by requiring three out of $n$ punctures to be at $0,1,\infty$, the one-form $\de$ becomes well-defined
in the tangent space to ${\CalM}_{0,n}$.

\subsection{The $SL_2$-opers in genus one}

We can also describe quite explicitly the space of opers on an elliptic curve $E_{\tau}$ with regular singularities at the points $x_{1}, \ldots , x_{n} \in E_{\tau}$:
\begin{equation}
\label{eq:ellop}
\begin{split}
{\CalD} = - {\p}_{z}^2 + T(z) \qquad\qquad\qquad\qquad \\
T(z) = u + \sum_{a=1}^{n}  
{\Delta}_{a} {\wp}(z - x_{a}) + {\ep}_{a} {\zeta}(z - x_{a}) \\
\end{split}
\end{equation}
where $\sum_{a} {\ep}_{a} = 0$, $u$ is a constant, and we used the Weierstrass ${\zeta}$ and ${\wp} = {\zeta}'$ functions. The correspondence ${\p}_{{\tau}_{k}} \leftrightarrow H_k$ is now represented by the one-form
\begin{equation}
{\de} = \sum_{a} {\ep}_{a}dx_{a} + u d{\tau}
\label{eq:corrsgo}
\end{equation}

\subsection{The opers on degenerate curves}

For practical purposes it is useful to study the regular opers on a smooth curve with punctures whose complex structure approaches  that of a maximally degenerate stable curve. As we discussed above, such maximal degeneration can be encoded in a trivalent graph $\Gamma$  with $n$ tails and $g$ loops. 
Each internal vertex corresponds to a three-holed sphere, 
each internal edge corresponds to a double point (a pinched handle). A complex structure close to the maximally degenerate one, corresponding to $\Gamma$, is parametrized by assigning the complex numbers $q_e$,
$|q_{e}| \ll 1$ to the internal edges. 

The oper $\CalD$ on $\Sigma$ which is close to $\Gamma$ can be approximated by the following data:
the hypergeometric oper on every trivalent vertex:
\begin{equation}
\label{eq:hypop}
T(z_v) = \frac{{\Delta}_{1}}{z_v^2} + \frac{{\Delta}_{2}}{(z_v-1)^2} + \frac{{\Delta}_{3} - {\Delta}_{1} - {\Delta}_{2}}{z_v(z_v-1)}
\end{equation}
where $z_v$ is a coordinate on the three-punctured sphere corresponding to the vertex $v$; 
and the glueing of the local coordinates across the edges:
\begin{equation}
z_{v_{1}(e)} z_{v_{2}(e)} = q_{e}
\label{eq:glue}
\end{equation}
where we assume that on the two spheres $v_1 (e)$ and
$v_2
(e)$ the double point corresponding to $e$ is at the points $z = 0$ on each component. 
Given the internal vertex $v$ let $e_1, e_2, e_3$ be the three emanating edges. Let them correspond to the points
$z_v = 0, z_v =1 , z_v = {\infty}$, respectively. 
Then, 
\begin{equation}
{\Delta}_{k} =  {\eta}_{k}  \left( {\eta}_{k}  - 1 \right)
\label{eq:alde}
\end{equation} 
where
${\eta}_{k} = \frac{{\al}_{e_k}}{2\pi i} + {\de}_k$ if $e_k$ is an internal edge, and ${\eta}_k = {\nu}_i$ if $e_k$ is the $i$'th tail, 
${\delta}_{1,3} = \frac{1}{2}$, ${\delta}_{2} = 0$. This shift is due to the fact that ${\CalD}$ acts on the $(-1/2)$-differentials 
$$
 {\Psi}_{v} = {\psi}(z_{v}) dz_{v}^{-\frac{1}{2}}
 $$

\subsection{Comparison with the four dimensional gauge theory}

In the perturbative limit, where the Riemann surface $\Sigma$ approaches one of the maximally degenerate complex structures ${\tau} \to {\tau}_{*}$ the ${\CalZ}$-function simplifies, and (\ref{eq:wfromz}) can be calculated rather explicitly \cite{NP:2011}:
\begin{equation}
\label{eq:wpert}
\begin{split}
{\CalW}_{{\CalO}_{\tau}}({\al}, {\n}) = \sum_{e} \left( \frac{{\tau}_{e}}{2} {\al}_{e}^{2} + {\Upsilon}(2{\al}_e ) \right)\\
\qquad\qquad - \sum_{v} {\Upsilon} (\sum_{e \ni v} \pm
{\al}_{e} ) \\
\end{split}
\end{equation}
where ${\Upsilon}$ is a special function, whose derivative gives the logarithm of the ${\Gamma}$-function:
\begin{equation}
\begin{split}
{\Upsilon}^{\prime}(x) = {\rm log} \left( \frac{{\Gamma}\left(x + {\ve} \right)}{{\Gamma}\left( - x + {\ve}\right)} \right) \qquad\qquad \\
{\Upsilon}(x) = \frac{d}{ds}\Biggr\vert_{s=0} \frac{{\mu}^{s}}{{\Gamma}(s)} \int_{0}^{\infty} \frac{dt}{t^2} \frac{t^{s}}{1-e^{-t{\ve}}} e^{-tx}
\end{split}
\label{eq:upgam}
\end{equation}
In order to compare this gauge theory result with the generating function of the variety of opers corresponding to our system of Darboux coordinates $({\al}, {\be})$ we approximate the oper on the nearly degenerate curve by the collection of the hypergeometric opers on the three-holed spheres corresponding to the internal vertices $v$
of the degeneration graph $\Gamma$, insert the transition matrices 
\begin{equation}
\label{eq:trmat}
\left(
\begin{matrix}
q_{e}^{{\al}_{e}} & 0 \\
0 & q_{e}^{-{\al}_{e}} \end{matrix} \right)
\end{equation}
at the double points, which correspond to the edges $e$,
and use the standard transition matrices for the solutions
of the hypergeometric equation, to compute the monodromy matrices ${\gm}_{1}, {\gm}_{2}, {\gm}_{3}, {\gm}_{4}$ for each genus zero edge, and the matrices
${\gm}, {\hm}$ for each genus one edge. 

One can also include the $e^{2\pi i {\tau}_{e}}$-corrections by the usual quantum mechanical perturbation theory and verify the agreement with the instanton calculations on the gauge theory side. We have performed these checks for low instanton numbers for the two basic cases: $SU(2)$ $N_f=4$ theory (which corresponds to ${\CalM}_{0,4}$) and for the $SU(2)$ ${\CalN}=2^{*}$ theory (which corresponds to ${\CalM}_{1,1}$). 

\subsection{The topological brane}

The study of the separation of variables proposed by E.~Sklyanin \cite{Sklyanin:1988} for the quantum Gaudin system, which is essentially the genus zero case of the Hitchin system, suggests the following definition of this second brane. We assume that the ${\nu}_{k}$ parameters are generic and not half-integral. For the connection of the separation of variables to the geometric Langlands program and related issues
see  \cite{Frenkel:1995ko}.

Then we define a subvariety ${\CalL}_{\gamma}$ for any pair ${\ga} = ({\Gamma}, {\sl or})$ which consists of the degeneration graph ${\Gamma}$ together with the choice of orientation {\sl or}. For the oriented edge $e$ we define the source $s(e)$ and the target $t(e)$ vertices in the obvious way.

The definition of ${\CalL}_{\gamma}$ is the following: 
{\em For every internal genus zero egde $e$ we require 
${\al}_{e}$ to be equal to the sum of 
 $\pm {\al}_e$'s or $\pm {\nu}_k$'s (the sign depends on the orientation) corresponding to the two other edges 
which enter $s(e)$}
$$
{\al}_{e} = \sum_{e^{\prime}, t(e^{\prime}) = s(e)} 
{\eta}_{e^{\prime}} - \sum_{e^{\prime}, s(e^{\prime}) = s(e)} 
{\eta}_{e^{\prime}}
$$
where, as before ${\eta}_{e} = {\al}_{e} \pm {\de}_{e}$ if $e$ is an internal edge, and ${\eta}_{e} = {\nu}_{k}$ if $e$ is the $k$'th tail. 

In the coordinate patch ${\CalU}_{{\Gamma}'}$ the variety ${\CalL}_{\ga}$ is described by the generating function, which is a sum of dilogarithms, as follows from (\ref{eq:csvol}). 

This is to be contrasted with the results of 
\cite{Frenkel:1995ko}, where the Gaudin eigenproblem in the case of half-integer ${\nu}_{k}$'s is solved by the trivial monodromy opers, which means, in the language of this paper, that the second brane corresponds to the unipotent monodromy flat connections. 
 
\section{Further directions and discussion}

In this paper we have introduced a system of holomorphic Darboux coordinates ${\al}, {\be}$ on the moduli space
${\CalM}_{\Sigma}^{loc}$ of flat $SL_{2}({\bC})$-connections on a punctured Riemann surface with fixed conjugacy classes of the monodromies around the punctures. The main claim about this coordinate system is the identification of the generating function of the variety
${\CalO}_{\tau}$ of $SL_2$-opers with the effective twisted superpotential of the four dimensional $A_1$ type  theory corresponding to $\Sigma$, subject to the two dimensional $\Omega$-deformation. 

We also expressed the Yang-Yang
function of the quantum Hitchin system, using our coordinate system, as a difference of the generating functions of the variety of opers ${\CalO}_{\tau}$, and the second Lagrangian submanifold ${\CalL}_{\ga}$, which determines the space of states of the quantum Hitchin system. We presented a proposal for the construction of 
${\CalL}_{\ga}$ in the genus zero case. As explained in \cite{Nekrasov:2009uh}, \cite{Nekrasov:2009zz}, \cite{Nekrasov:2009ui} the Yang-Yang function we are talking about here is different from the Yang-Yang functions of the finite dimensional Gaudin model or spin chains, which can be derived by taking a critical level limit of the free field representation of the current algebra conformal blocks \cite{Gerasimov:1990fi}, \cite{Varchenko:1989}, as in \cite{Feigin:1994in}, \cite{Reshetikhin:1994}, \cite{Felder:1995ge}.

The generating function ${\CalW}_{{\CalO}_{\tau}}$ also has other applications. For example, it can be identified with the classical conformal block of the Liouville theory \cite{ZZ:1995}, which makes the relation to the four dimensional gauge theory natural in view of the conjecture \cite{Alday:2009aq}. It also provides the "holomorphic part" of the classical Liouville action, which is discussed
in \cite{Polyakov:1987}, \cite{Alekseev:1988ce}, \cite{Takhtajan:1988}, \cite{Verlinde:1990h}, \cite{Aldrovandi:2000}. 

Since the variety of opers is well-understood for all Lie groups $G$, it is urgent to generalize our Darboux coordinates for the case of general $G$. This would allow us to compute the effective twisted superpotentials of the exotic theories which do not have a Lagrangian description
(a general $A,D,E$ type $(0,2)$ theory compactified
on a Riemann surface and subject to the $\Omega$-deformation).

In order to characterize the conformal blocks of Liouville
and Toda conformal theories, using the \cite{Alday:2009aq} relation, we need to turn on the generic $\Omega$-deformation, with both ${\ve}_{1}, {\ve}_{2}$ non-vanishing. It was argued
that this would effectively quantize the algebras of holomorphic functions on ${\CalM}^{loc}_{\Sigma} (G)$ and 
${\CalM}^{loc}_{\Sigma} (^{L}G)$, with the deformation quantization parameters ${\hbar} = {\ve}_{2}/{\ve}_{1}$ and
$1/{\hbar} = {\ve}_{1}/{\ve}_{2}$, respectively. The ${\CalZ}$-function is then a vector in the representation of
${\CalA}_{G}^{\hbar} \times {\CalA}_{^{L}G}^{1/{\hbar}}$, which corresponds, quasiclassically, to ${\CalO}_{\tau}$. To find this vector seems like an extremely important problem. For the recent discussion of related problems see \cite{Chekhov:1999tn}, \cite{Alday:2009fs}, \cite{Teschner:2010qh}, \cite{Nekrasov:2010ka}. 

Finally, let us mention a couple more problems we left out of this short note. 

Our construction of the Darboux coordinates $({\al}, {\be})$ and the description of the variety of opers ${\CalO}_{\tau}$ must have an interesting quasiclassical limit, corresponding to the ${\ve} \to 0$ limit of the $\Omega$-background. In this limit one should recover the Seiberg-Witten geometry of the Hitchin system. 

Also, a large mass, weak coupling limit of the gauge theory, as in e.g. \cite{Gaiotto:2009ma}, leading to the asymptotically free gauge theory must have a counterpart in our construction. We believe extending our Darboux coordinates to the case of the moduli spaces of flat connections with irregular singularities will play an important role both in the analysis of the four dimensional gauge theory, and in the extension of the Langlands duality to the case of the wild ramification \cite{Witten:2007td}, as well as
in the applications to the integrable systems \cite{Feigin:2010ir}. 

\section{Acknowledgements}

We are grateful to V.~Fock, E.~Frenkel, A.~Gerasimov, Yu.~Neretin, F.~Smirnov for valuable discussions. Research of NN was partly supported by FASI RF 14.740.11.0347 
that of AR was partly supported by RFBR-09-02-00393, FASI RF 14.740.11.0347, RFBR-09-01-93106-NCNIL-a,
and RF Government Grant-11.G34.31.0023, that of SS by SFI grant 08/RFP/MTH1546 and by
ESF grant ITGP. Part of the research was done while AR visited IHES in the summer of 2010. NN also thanks Ilya Khrzhanovsky for the opportunity to carry out some of the research on the project at the "DAU" Institute in Kharkiv, Ukraine, and to D.~Kaledin and A.~Losev for discussions. 

The main conjecture about the generating function of the variety of opers was announced by one of the authors at the Simons Center workshop on "Perspectives, Open Problems \& Applications of Quantum Liouville Theory" (Mar 29-Apr 2, 2010). The results of this paper were also reported at various seminars and conferences, in particular, at the Institut Henri Poincare, at the conferences "Symmetry, Duality, and Cinema" (Jun 2010) and  "Advances in string theory, wall crossing,
and quaternion-K\"ahler geometry" (Aug 30 - Sept 3, 2010), at the Cargese Summer School "String Theory: Formal Developments and Applications" (Jun 21-Jul 3, 2010) at the IAS (Princeton) seminar (Oct 29, 2010), at the DESY
conference "From Sigma Models to Four-dimensional QFT", 
 (Nov 29 - Dec 3, 2010), at the Hebrew Univeristy of Jerusalem IAS 15th Midrasha Mathematicae on
�Derived Categories of Algebro-Geometric Origin and Integrable systems"
(Dec 19 - 24, 2010). We thank all the organizers and the audiences for the fruitful atmosphere and 
              stimulating questions.

\end{document}